\documentclass[]{elsart}

\usepackage{epsfig}
\usepackage{graphics}
\usepackage{graphicx}
\usepackage[centertags]{amsmath}
\usepackage{multicol}
\usepackage{captcont}
\usepackage{url}
\usepackage{subfigure}
\usepackage{algorithmic}
\usepackage{hyperref}

\newtheorem{finding}{Finding}

\pdfoutput=1

\begin{document}

\begin{frontmatter}

\title{Approximating Mexican highways \\ with slime mould}

\author{Andrew Adamatzky$^1$, Genaro J. Mart{\'i}nez$^{1,2}$}
\author{Sergio V. Chapa-Vergara$^3$, Ren\'e Asomoza-Palacio$^3$} 
\author{Christopher R. Stephens$^2$}

\address{$^1$~Unconventional Computing Centre, University of the West of England, Bristol BS16 1QY, United Kingdom \\
\{andrew.adamatzky, genaro.martinez\}@uwe.ac.uk \\
$^2$~Instituto de Ciencias Nucleares and Centro de Ciencias de la Complejidad, Universidad Nacional Aut\'onoma de M\'exico, M\'exico \\
$^3$~Ingenier{\'i}a El\'ectrica, Departamento de Computaci\'on, Centro de Investigaci\'on y de Estudios Avanzados del Instituto Polit\'ecnico Nacional, M\'exico} 
\date{}

\maketitle

\begin{abstract}

\noindent
Plasmodium of \emph{Physarum polycephalum} is a single cell visible by unaided eye. During its foraging behavior 
the cell spans spatially distributed sources of nutrients with a protoplasmic network. Geometrical structure of the 
protoplasmic networks allows the plasmodium to optimize transport of nutrients between remote parts of its body. Assuming
major Mexican cities are sources of nutrients how much structure of \emph{Physarum} protoplasmic network correspond to 
structure of Mexican Federal highway network? To find an answer undertook a series of  laboratory experiments with living  
\emph{Physarum polycephalum}. We represent geographical locations of major cities (nineteen locations) by oat flakes,
place a piece of plasmodium in Mexico city area, record the plasmodium's foraging behavior and extract topology of nutrient transport networks. Results of our experiments show that the protoplasmic network formed by \emph{Physarum} is isomorphic, subject to limitations imposed,  to a network of principle highways. Ideas and results of the paper may contribute towards future developments in bio-inspired road planning. 

\vspace{0.5cm}

\end{abstract}

\end{frontmatter}

\section{Introduction}

Plasmodium of \emph{Physarum polycephalum}\footnote{Order \emph{Physarales}, subclass \emph{Myxogastromycetidae}, class \emph{Myxomecetes}} is a single cell with many diploid nuclei. The plasmodium feeds on microbial creatures and microscopic 
food particles. When colonizing its habitat the plasmodium develops an optimal network of protoplasmic tubes or veins. 
The protoplasmic network is optimal in a sense that it maximizes number of food sources occupied,
minimizes time for nutrient transport between distant parts of the plasmodium's body, and maximizes area of substrate covered by sensorial activity.

A plasmodium is a single cell yet can be considered as large-scale collective of simple entities with distribute and massive-parallel sensing, computation and actuation. A sensing is parallel because the plasmodium can detect and determine position of many source of chemo-attractants, including nutrients, and also make decentralized sensing of environmental conditions, including humidity, temperature and illumination. An actuation is parallel because the plasmodium can propagate in several directions in parallel, the plasmodium can occupy and colonize many food sources at the same. With regards to parallel 
computation, the plasmodium is a wave-based massively-parallel reaction-diffusion chemical computer~\cite{adamatzky_2005,adamatzky_bz_trees,adamatzky_physarummachines}. A computation in the plasmodium is implemented by interacting bio-chemical and excitation waves~\cite{nakagaki_yamada_1999}, redistribution of electrical charges on plasmodium's membrane~\cite{achenbach_1981} and spatio-temporal dynamics of  mechanical waves~\cite{nakagaki_yamada_1999}. 
Experimental proofs of \emph{P. polycephalum} computational abilities include approximation 
of shortest path~\cite{nakagaki_2001a} and hierarchies of planar proximity graphs~\cite{adamatzky_ppl_2009},
computation of plane tessellations~\cite{shirakawa}, implementation of primitive memory~\cite{saigusa}, 
execution of basic logical computing schemes~\cite{tsuda_2004}, control of robot navigation~\cite{tsuda_2007}, 
and natural implementation of spatial logic and process algebra~\cite{schumann_adamatzky_2009}. See overview of Physarum-based computers in \cite{adamatzky_physarummachines}. 

Approximation of shortest or even computation of a transportation network have been already a hot application for 
unconventional computing scientists. Nature-inspired computing paradigms and experimental implementations were successfully applied 
to calculation of a minimal-distance path between two given points in a space or a road network. Thus, a 
shortest-path problem is solved in experimental reaction-diffusion chemical systems~\cite{adamatzky_2005},
gas-discharge analog systems~\cite{reyes_2002}, spatially extended crystallization systems~\cite{adamatzky_hotice},
formation of fungi mycelian networks~\cite{jarret_2006}, and using computer and mathematical models of collective insects~\cite{dorigo_2004} and \emph{Physarum polycephalum}~\cite{tero_2006}.  
Previously~\cite{adamatzky_UC07} we have evaluated a road-modeling potential of \emph{P. polycephalum}, 
however no conclusive results were presented back in 2007.  A step forward biological-approximation, or evaluation, of man-made 
road networks was done in our previous paper on approximation of United Kingdom motorways by plasmodium of
\emph{P. polycephalum}~\cite{adamatzky_jones_2009}. Results obtained in~\cite{adamatzky_jones_2009} were refreshing yet convincing. In general, we shown that transportation links constructed by Physarum match man-made motorways, there are however subtle differences between slime mould and motorways, see details in~\cite{adamatzky_physarummachines}. The experimental laboratory approach to slime-mould-based approximation of road networks was tested on just one country (United Kingdom) so far. To make
this bio-inspired approach truly universal we must study how the plasmodium will behave on different setups of cities and different shapes of countries. In present paper we have chosen Mexico as a test country for plasmodium approximation because it is the most populated Spanish-speaking country with diverse landscapes, sudden north-south changes in land size, highest level of public transport in Latin America, and high-concentration pollutants in central part of the country. These features are not unique but enhanced and, what it is most important, they provide very good test-bed for studies of nature-inspired approaches to road planning. 

Mexico ranks 7th in highways length with 6,335 kilometers (3,935 miles). Followed by (1)~United States with 91,541 km (56,859 mi), (2)~China with 24,474 km (15,202 mi), (3)~Germany with 11,400 km (7,081 mi), (4)~France with 10,300 km (6,398 mi), (5)~Spain with 9,063 km (5,629 mi), and Italy with 8,957 km (5,563 mi).\footnote{\url{http://www.publicpurpose.com/hwy-worldmotorway.htm}} Political division constitute 31 states and one Federal District (DF or Mexico city) that centralize the main activities in the whole republic, for this reason main highways at least connect one time Mexico city from any place. With a total population of 103,263,388 (2005 estimation) on a territorial extension of 1,964,375 km$^2$ (758,450 mi$^2$).\footnote{National Institute of Statistical and Geographic (INEGI Spanish abbreviation) \url{http://inegi.org.mx/}} Mexico has a relevant and strategical geographical position because it is a natural land transit between North America (mainly United States and Canada) with Mexico self, Central and South America, i.e., all Latin America countries. This contributes towards essential components of economical structure for the United States in transporting prime materials, natural resources, imports and exports, and labor migration.
For this reason years ago extensive highways were built from Central America across Mexico to the United States, 
and from coast to coast. 

The paper is structured as follows. In Sect.~\ref{methods} we describe experimental setup and software analysis. Experimental laboratory results on Physarum approximation of transport links are presented in Sect.~\ref{results}. Analysis of Physarum and man-made highways in terms of planar proximity graphs is provided in Sect.~\ref{proximitygraphs}.
Directions of future studies are outlined in Sect.~\ref{discussion}.

\section{Methods}
\label{methods}

Plasmodium of \emph{ P. polycephalum} is cultivated in plastic container, on paper kitchen towels sprinkled with still drinking water and fed with oat flakes\footnote{Asda's Smart Price Porridge Oats}. For experiments we use $120 \times 120$~mm polystyrene square and round Petri dishes. We use 2\% agar gel (Select agar, Sigma Aldrich). Agar plates are cut in a shape of Mexico.

\begin{figure}[!tbp]
\centering
\subfigure[]{\includegraphics[width=0.75\textwidth]{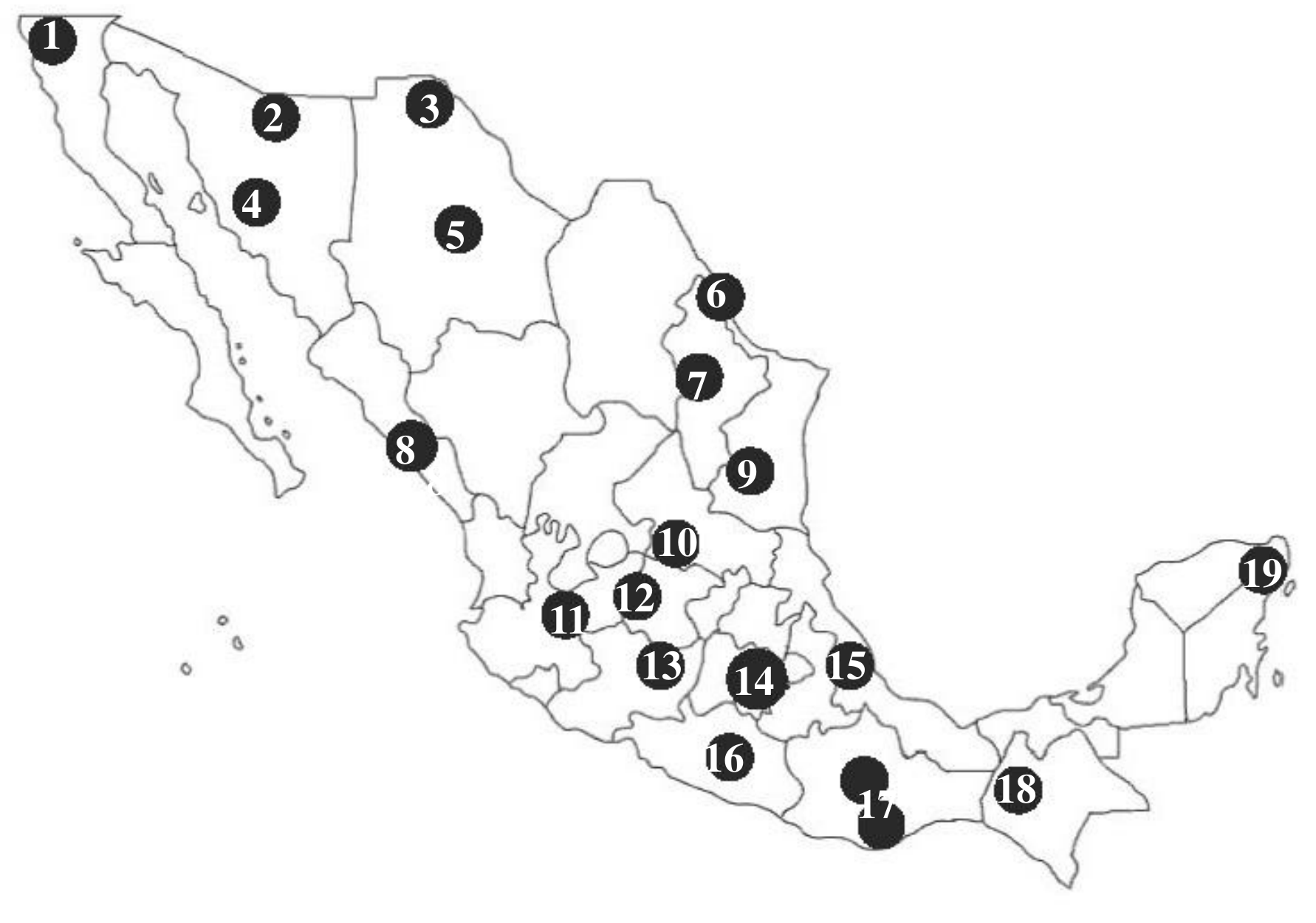}}
\subfigure[]{\includegraphics[width=0.49\textwidth]{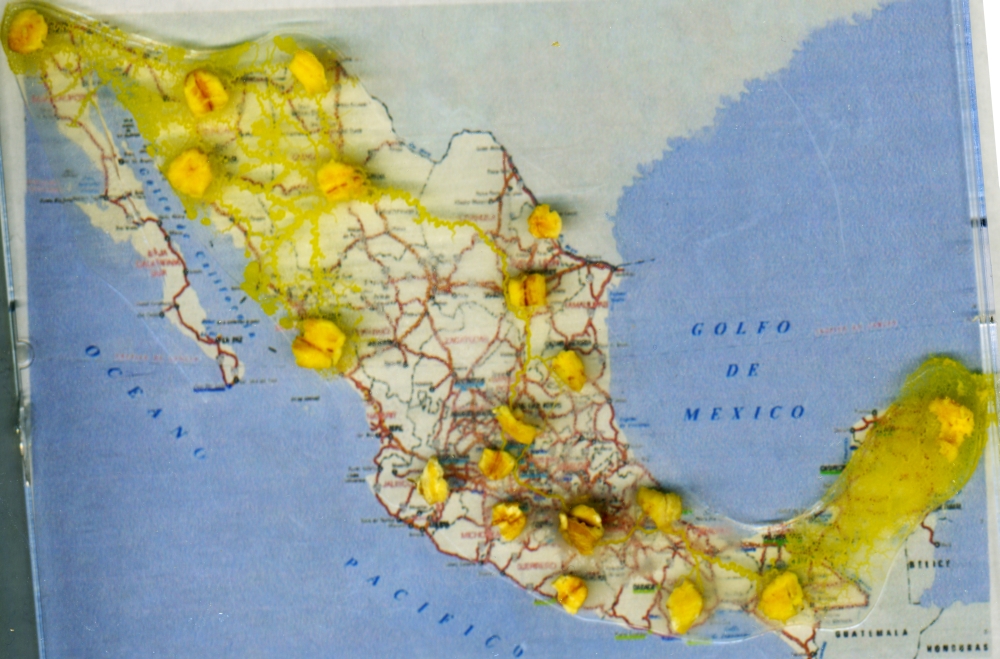}}
\subfigure[]{\includegraphics[width=0.49\textwidth]{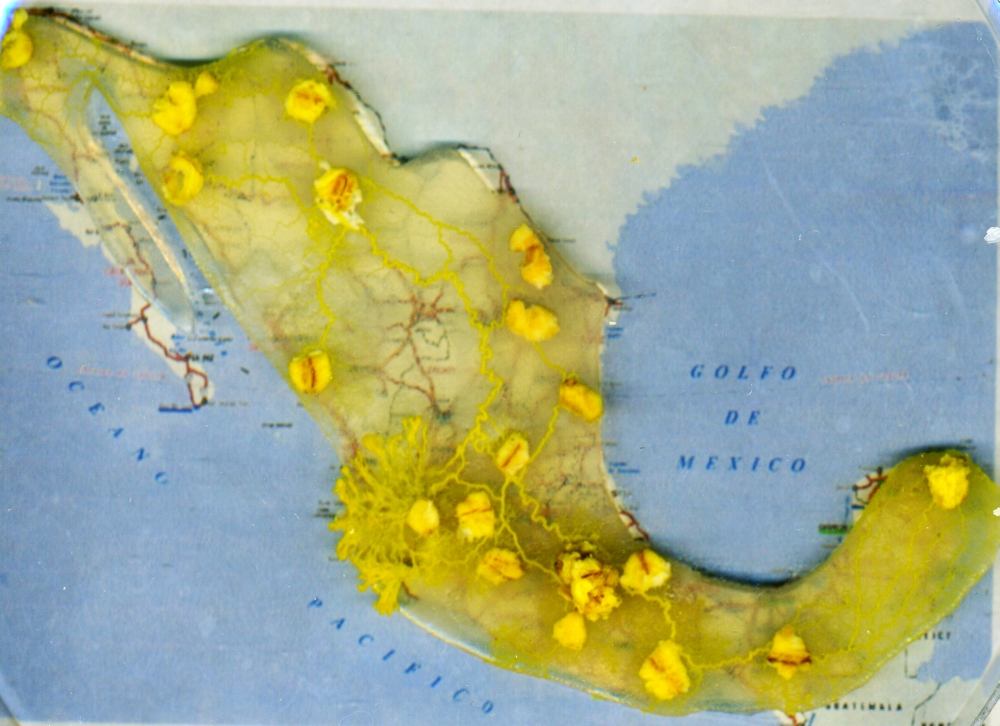}}

\caption{Experimental basics. 
(a)~Contour map of Mexico with 19 sources of nutrients indicated. (b)--(c)~Snapshots of two typical setups: urban areas are represented by oat flakes, plasmodium is inoculated in Mexico city, the plasmodium spans oat flakes by protoplasmic transport network. Note that configurations of protoplasmic tubes are different in these two experiments.}
\label{urbanareas}
\end{figure}

\begin{figure}[!tbp]
\centering
\includegraphics[width=1.0\textwidth]{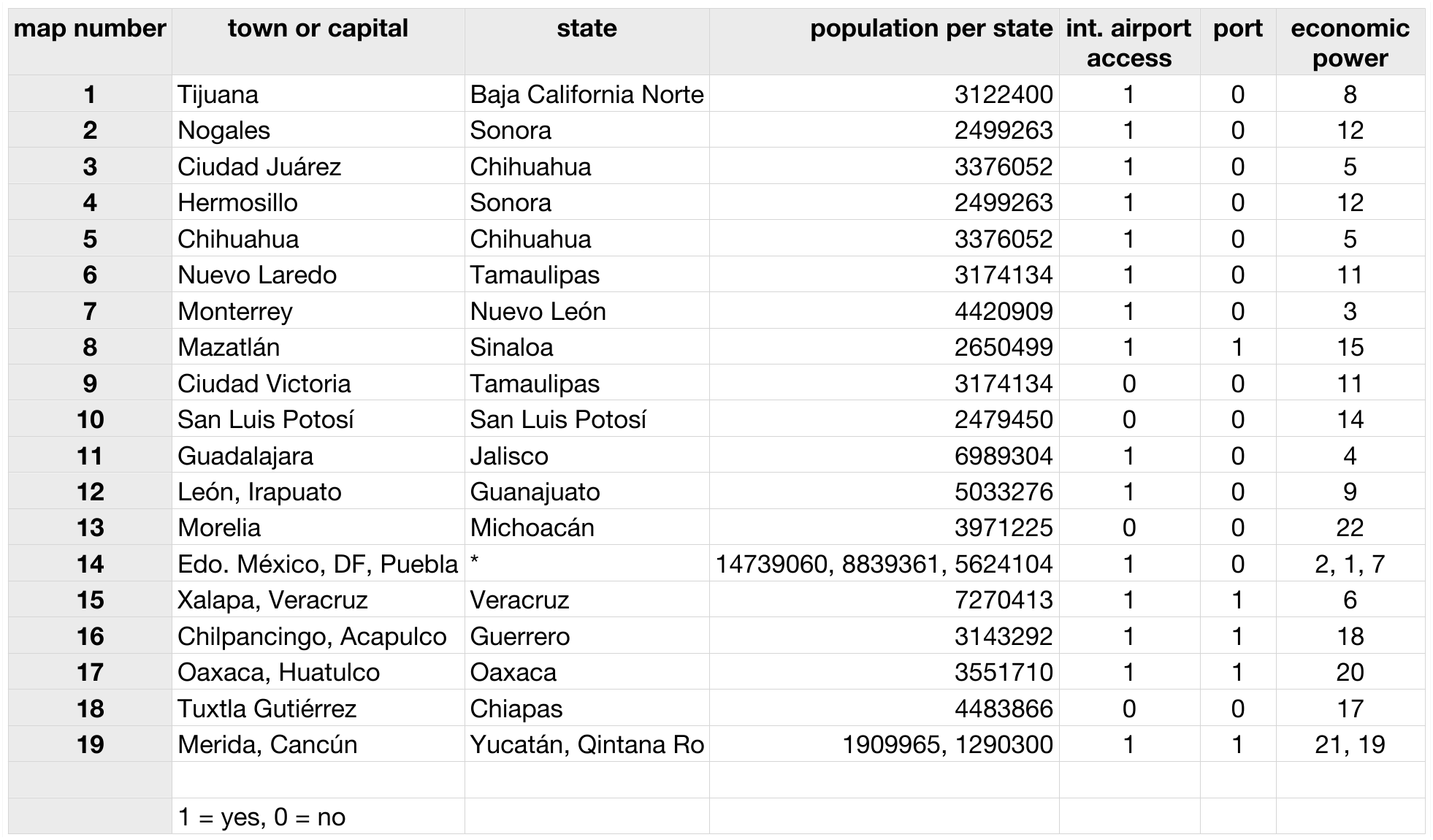}
\caption{General data regions $\mathbf U$ (Fig.~\ref{urbanareas}a), including main town and state, 
population, access to international air and sea ports, and economic potential (INEGI source 2010).}
\label{generalData}
\end{figure}

We selected 19 most populous urban areas, based on a selection of relevant features, including 
economic impact, sea and air ports, and tourist attractions. Details of the areas are provided in 
Fig.~\ref{generalData}.


Further we refer to the urban regions as $\mathbf U$. These region of $\mathbf U$ are projected onto gel and oat flakes,  of size approximately  matching size and shape of the regions, are placed in the positions of the regions (Fig.~\ref{urbanareas}b). At the beginning of each experiment a piece of plasmodium, usually already attached to an oat flake, is placed in the region corresponding to Edo. M\'{e}xico, DF, Puebla (region 14 in Fig.~\ref{urbanareas}a). 

The Petri dishes with plasmodium are kept in darkness, at temperature 22-25~C$^{\text o}$, except for observation and image recording. We undertook 26 experiments. Periodically the dishes are scanned using Epson Perfection 4490 scanner. Scanned images of dishes are enhanced for higher visibility, saturation increased to 55, and contrast to 40. 

To ease readability of experimental images we provide a binary version of each image. The binarization
is done as follows. Each pixel of a color image is assigned black color if red $R$ and green $G$ components of its
RGB color exceed some specified thresholds, $R > \theta_R$, $G > \theta_G$ and blue component $B$ does 
not exceed some threshold value $B < \theta_B$; otherwise, the pixel is assigned white color 
(exact values of the thresholds are indicated in the figure captions as $\Theta=(\theta_R, \theta_G, \theta_B)$.

\section{Transport links via foraging}
\label{results}

\begin{figure}[!tbp]
\centering
\subfigure[$t=$12~h]{\includegraphics[width=0.4\textwidth]{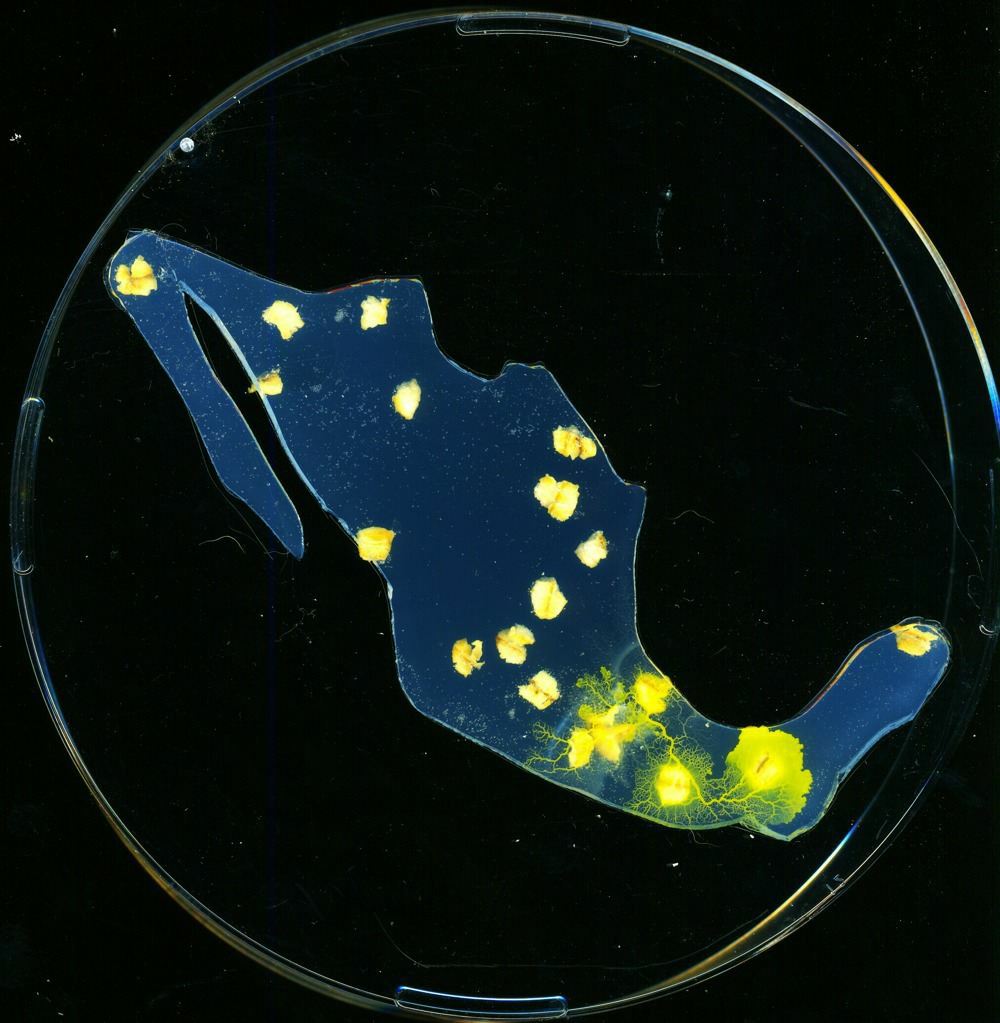}}
\subfigure[$t=$24~h]{\includegraphics[width=0.4\textwidth]{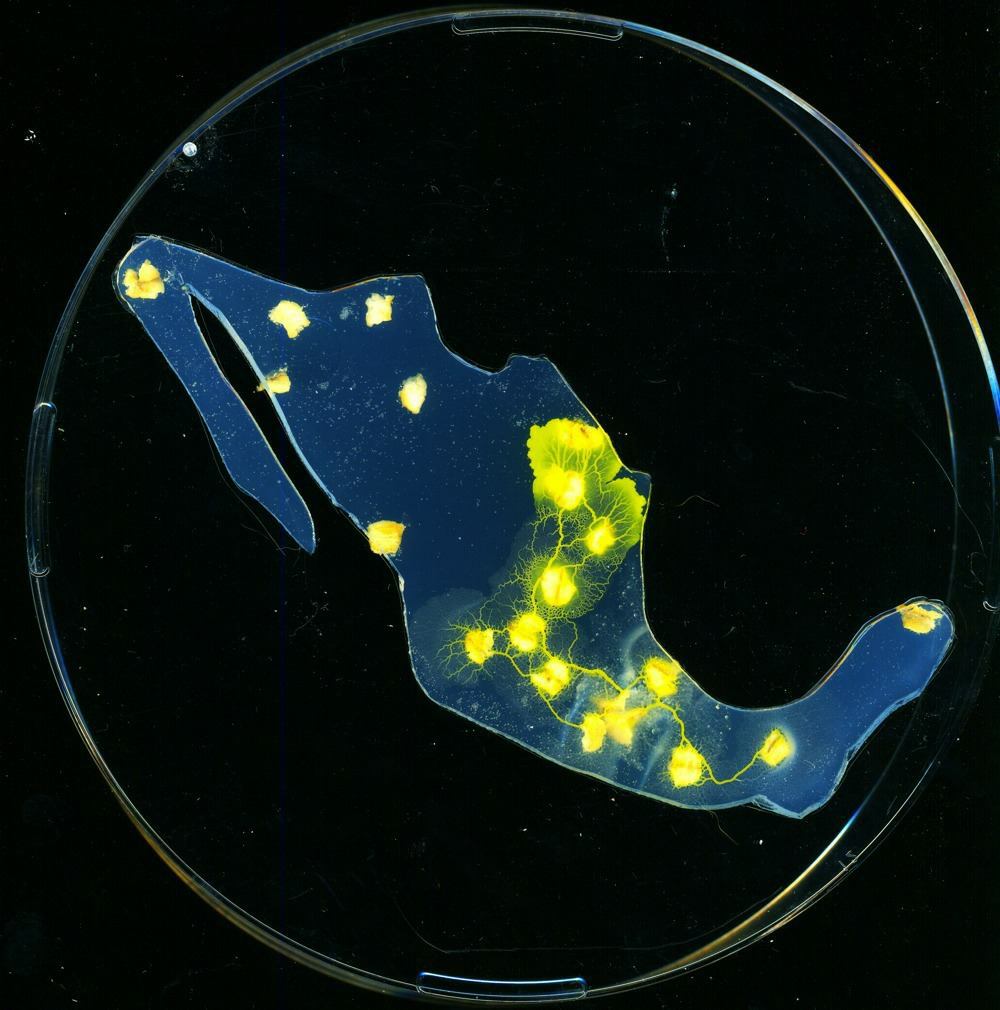}}
\subfigure[$t=$38~h]{\includegraphics[width=0.4\textwidth]{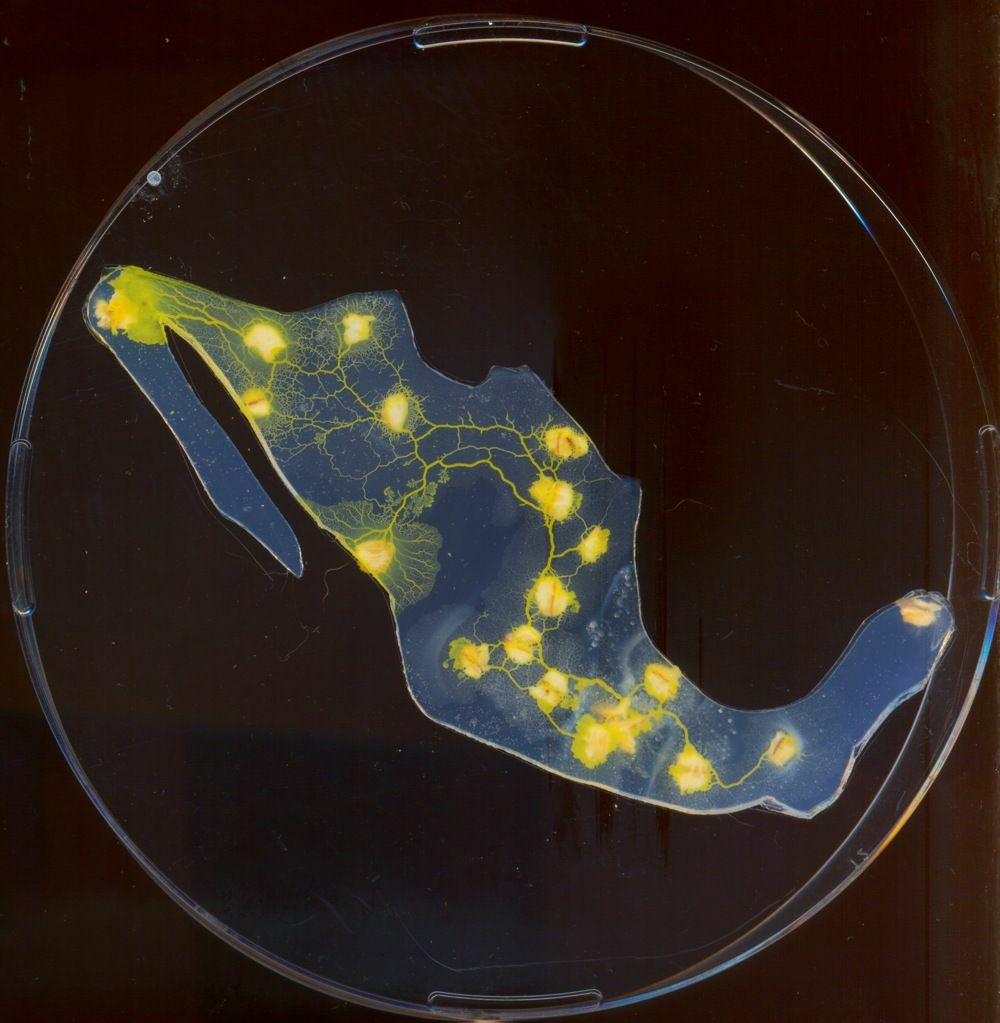}}
\subfigure[$t=$52~h]{\includegraphics[width=0.4\textwidth]{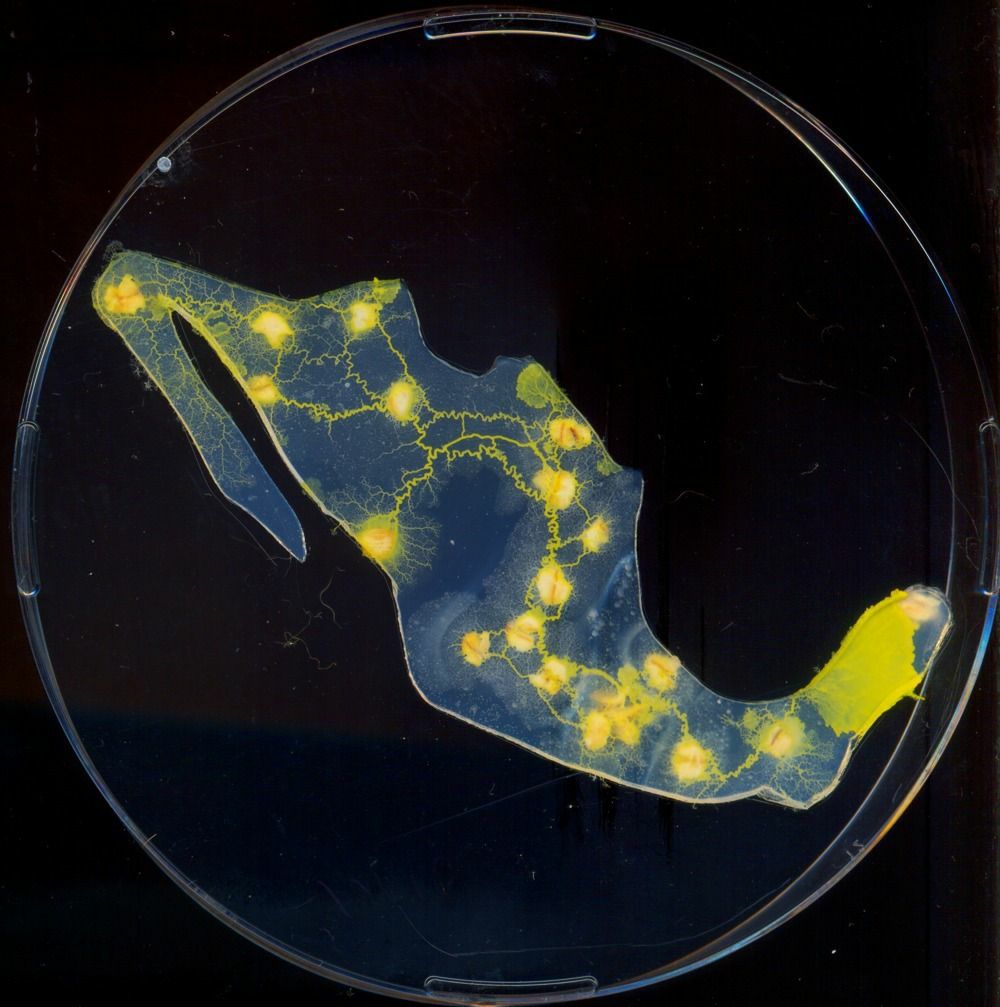}}
\subfigure[$t=$61~h]{\includegraphics[width=0.4\textwidth]{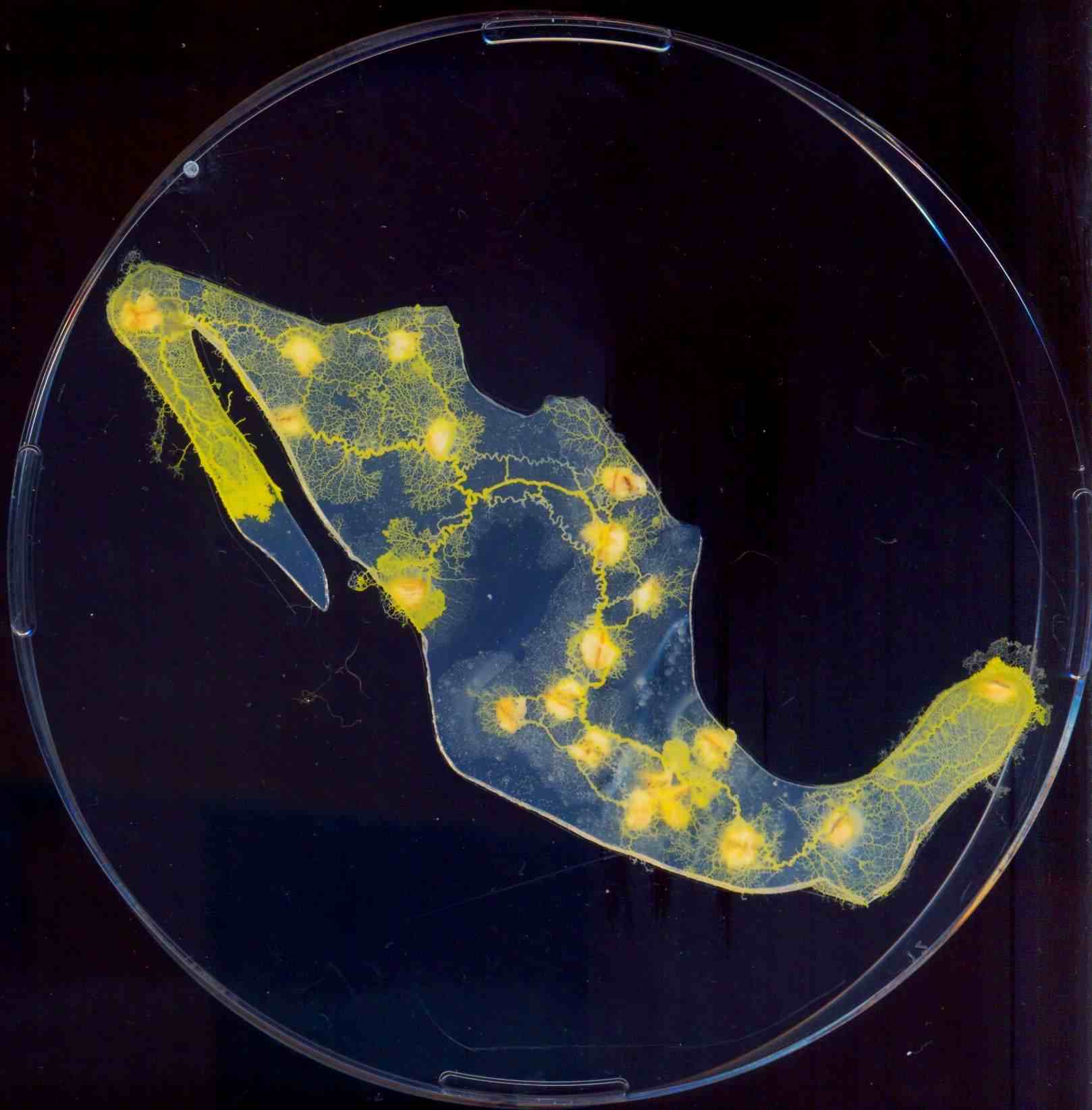}}
\captcont{Typical plasmodium development:
(a)--(e)~scanned image of experimental Petri dish. Time elapsed from inoculation is shown in the sub-figure captions.
(f)--(j)~binarized images, $\Theta=(100,100,100)$. 
}
\label{am9}
\end{figure}

\begin{figure}[!tbp]
\centering
\subfigure[$t=$12~h]{\includegraphics[width=0.4\textwidth]{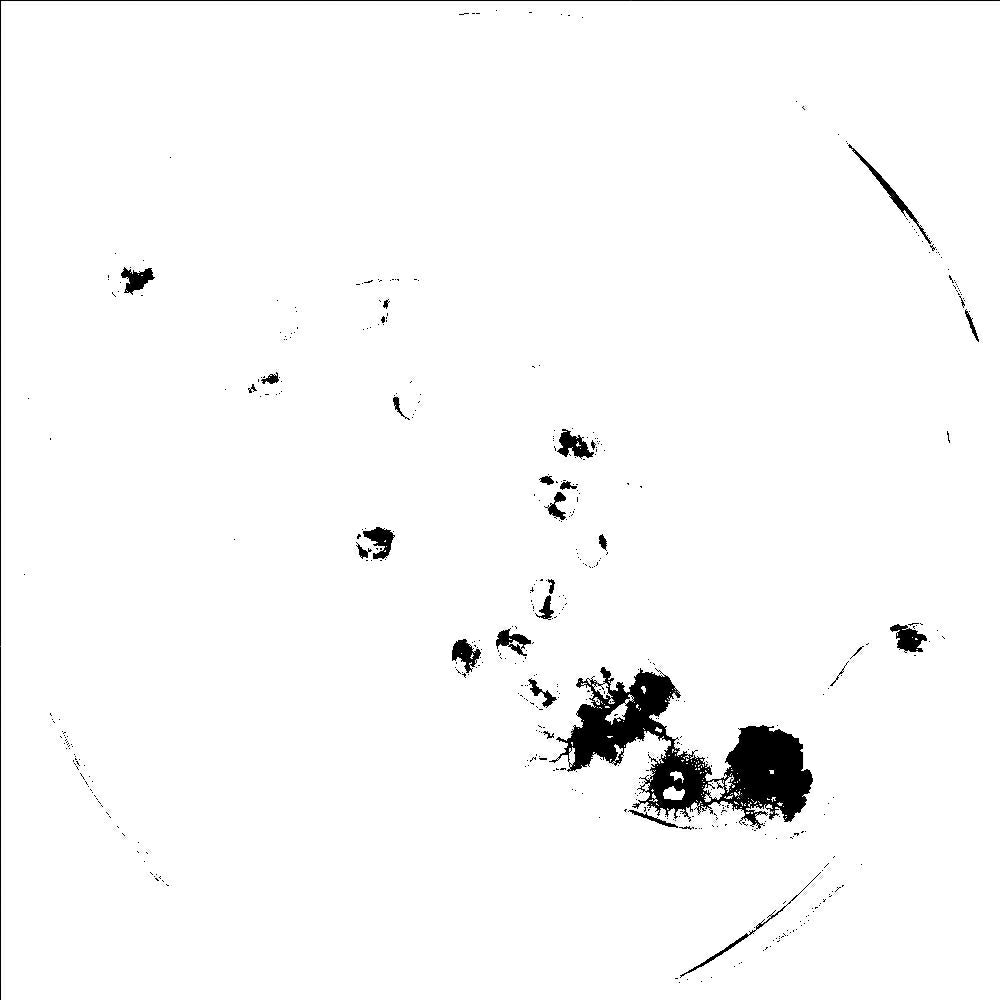}}
\subfigure[$t=$24~h]{\includegraphics[width=0.4\textwidth]{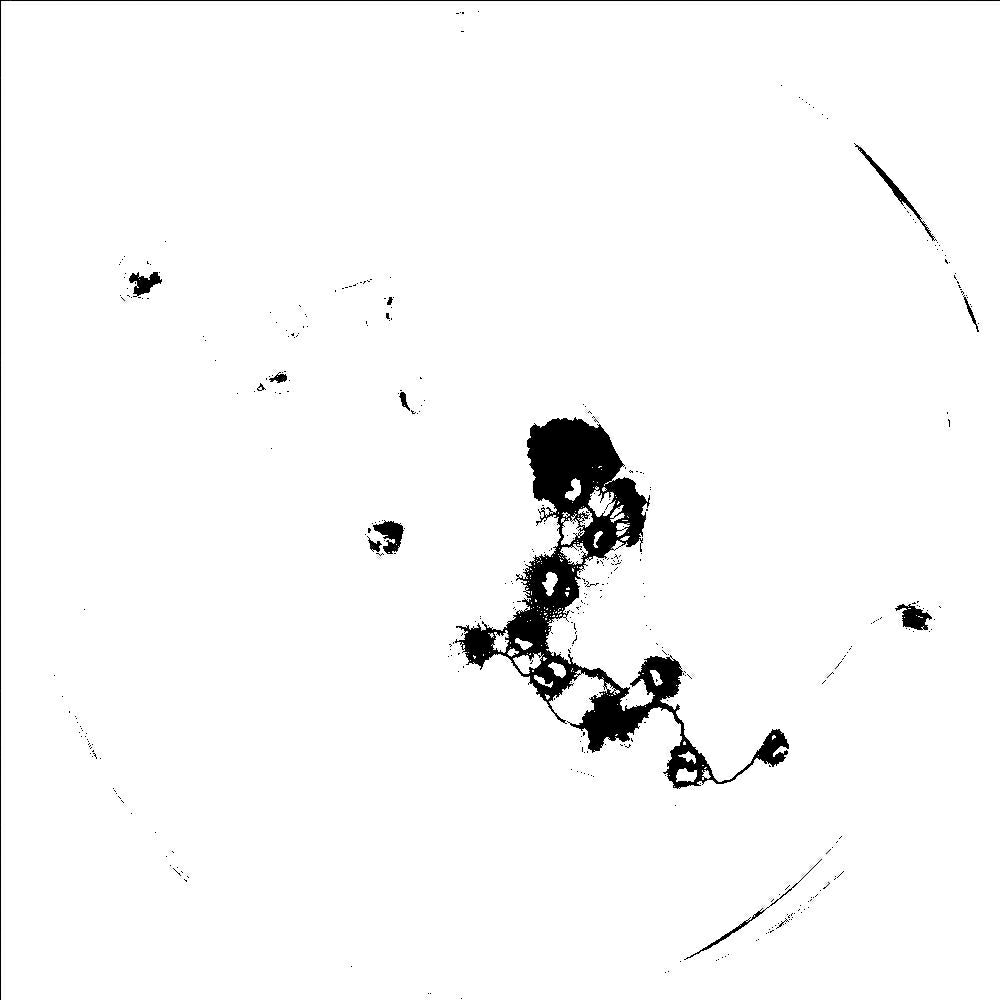}}
\subfigure[$t=$38~h]{\includegraphics[width=0.4\textwidth]{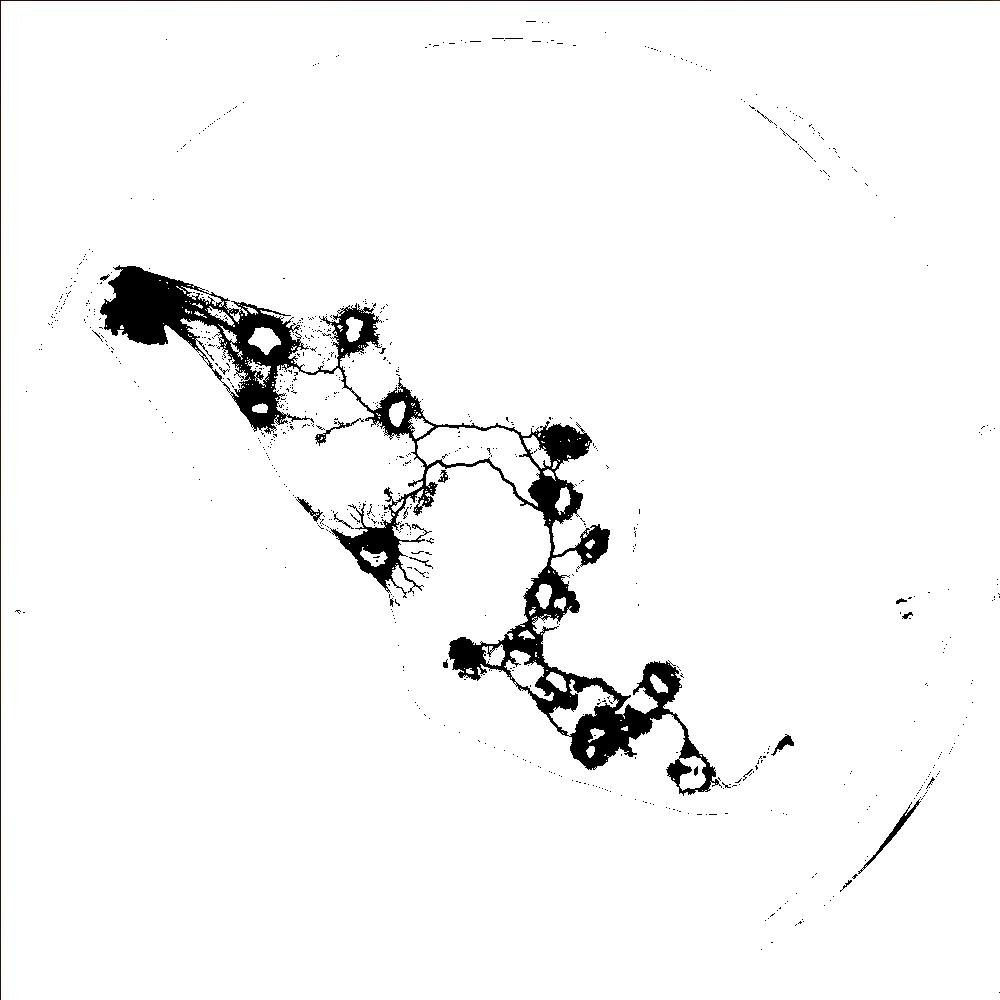}}
\subfigure[$t=$52~h]{\includegraphics[width=0.4\textwidth]{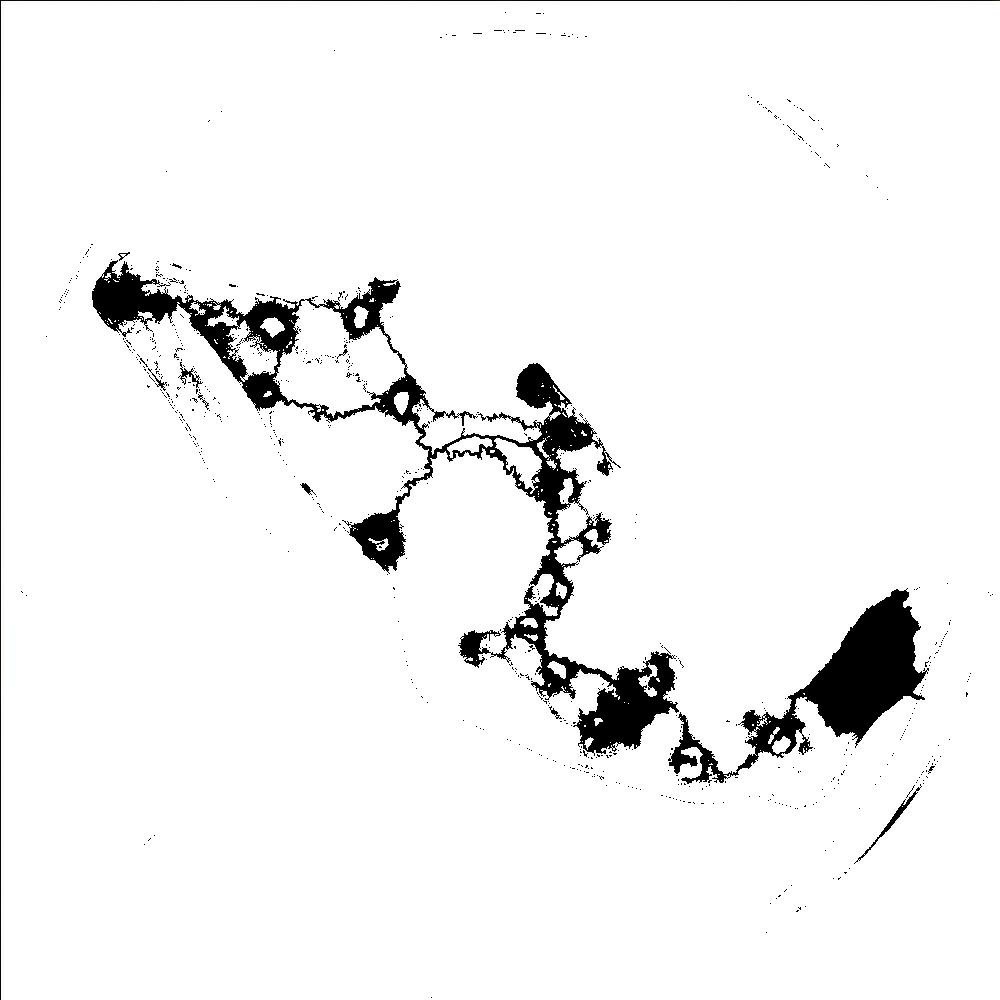}}
\subfigure[$t=$61~h]{\includegraphics[width=0.4\textwidth]{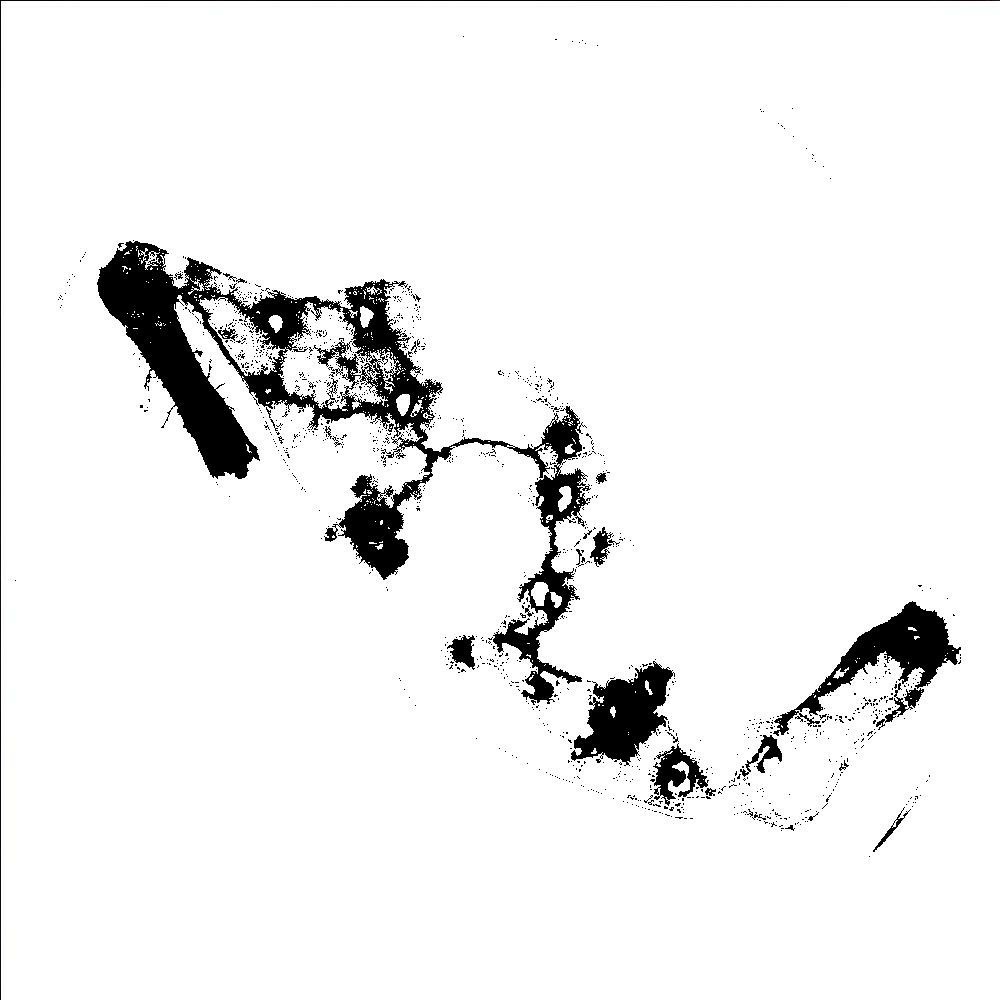}}
\caption{Continued.}
\label{am9}
\end{figure}

Snapshots of plasmodium foraging patterns during typical experiment are shown in Fig.~\ref{am9}. Initially 
a piece of plasmodium is placed onto an oat flake representing Mexico City. The plasmodium grows and propagates
in south-easterly direction  occupying Oaxaca-Huatulco region and then traveling towards Tuxtla Guti\'errez 
region (Fig.~\ref{am9}af). The plasmodium reaches Tuxtla Guti\'errez region usually in 12 hours after inoculation (real time distance is 11 hours). 
At the same time the plasmodium colonizes Xalapa-Veracruz and Chilpancingo-Acapulco regions and moves 
north-north-west. In a normal situation (vigorous plasmodium, fresh agar gel, air not contaminated by bacteria), 
south-east and north-north-west developments occurs simultaneously. In 12-14 hours after inoculation the plasmodium can 
reach as far as Monterrey and Nuevo Laredo regions (Fig.~\ref{am9}bg).  The plasmodium colonizes almost the whole Mexico 
in 38 hours, spanning major urban region from Tijuana in the north-west to Tuxtla Guti\'errez in the south-east (Fig.~\ref{am9}ch).
The colonization is truly completed in 52 hours, when an oat flake representing Merida-Canc\'{u}n is covered by plasmodium 
mass (Fig.~\ref{am9}di). Despite spanning all cities marked by oat flakes the plasmodium continues exploring the territory outlined
by agar gel and propagates along Baja California Peninsula (Fig.~\ref{am9}ej). There are no cities represented by food sources, therefore the plasmodium retreats from the Peninsula.

\begin{figure}[!tbp]
\centering
\subfigure[$t=$12~h]{\includegraphics[width=0.4\textwidth]{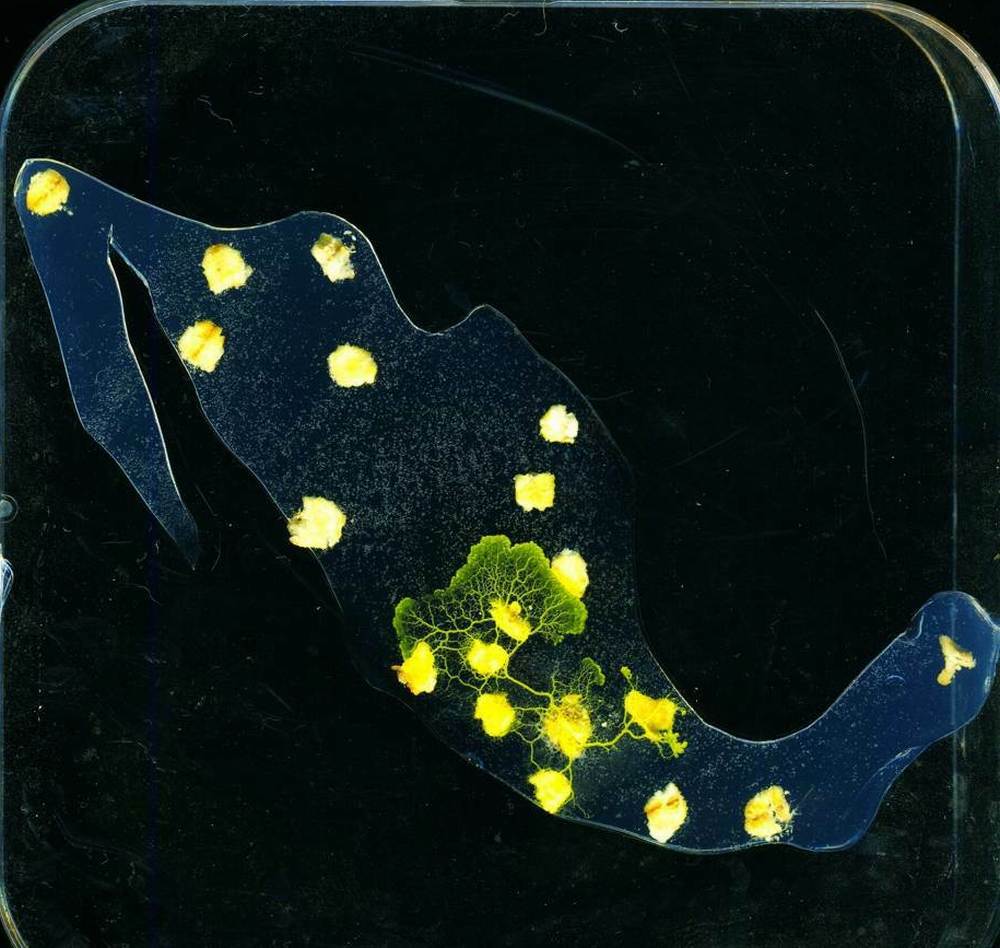}}
\subfigure[$t=$24~h]{\includegraphics[width=0.4\textwidth]{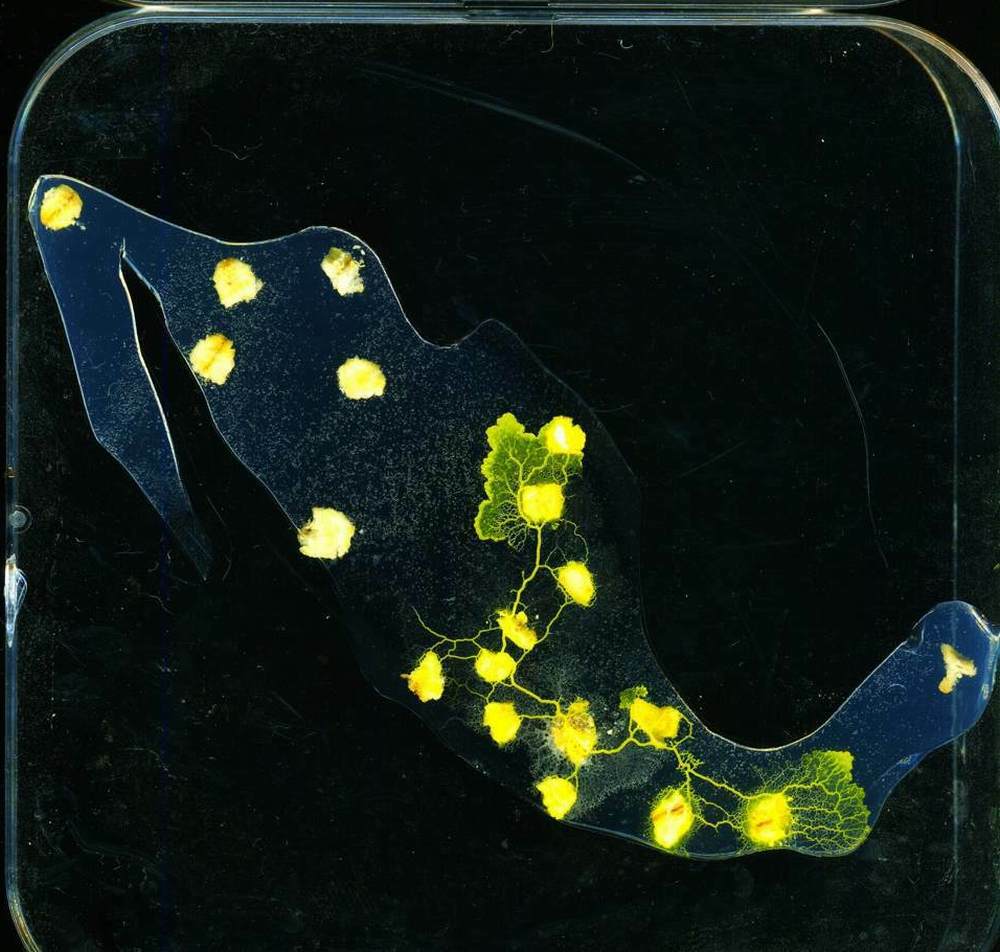}}
\subfigure[$t=$35~h]{\includegraphics[width=0.4\textwidth]{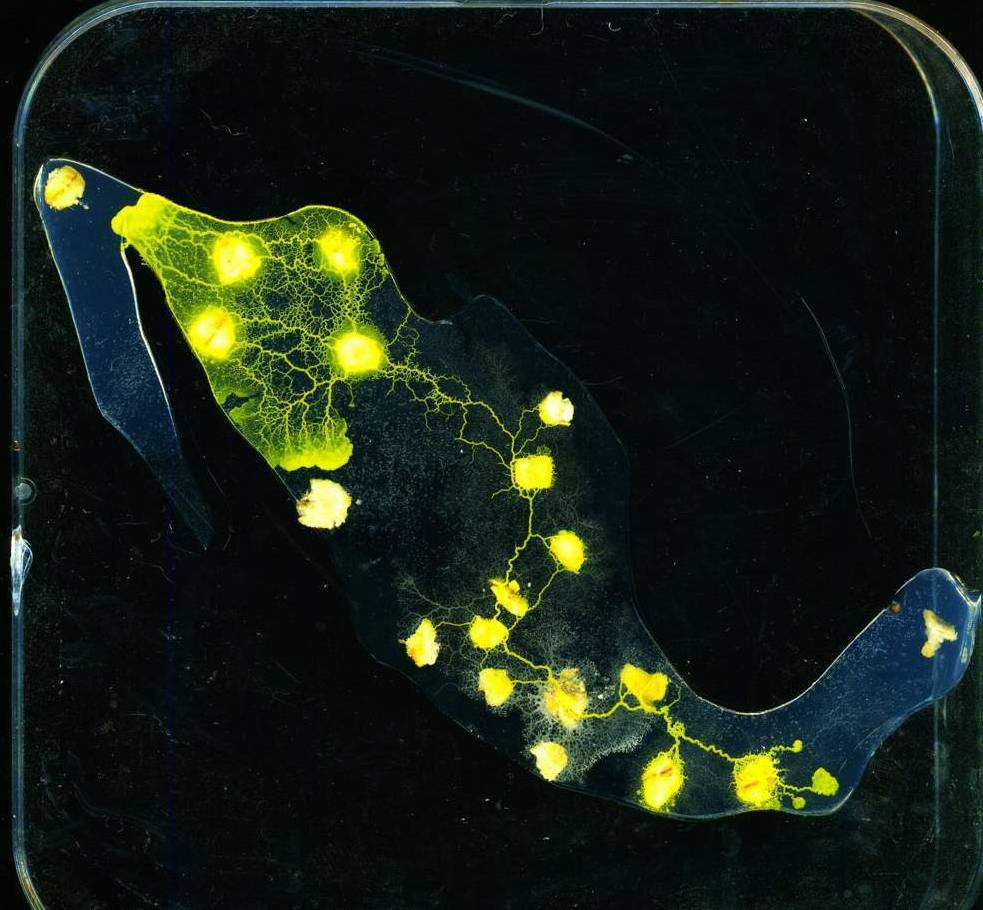}}
\subfigure[$t=$47~h]{\includegraphics[width=0.4\textwidth]{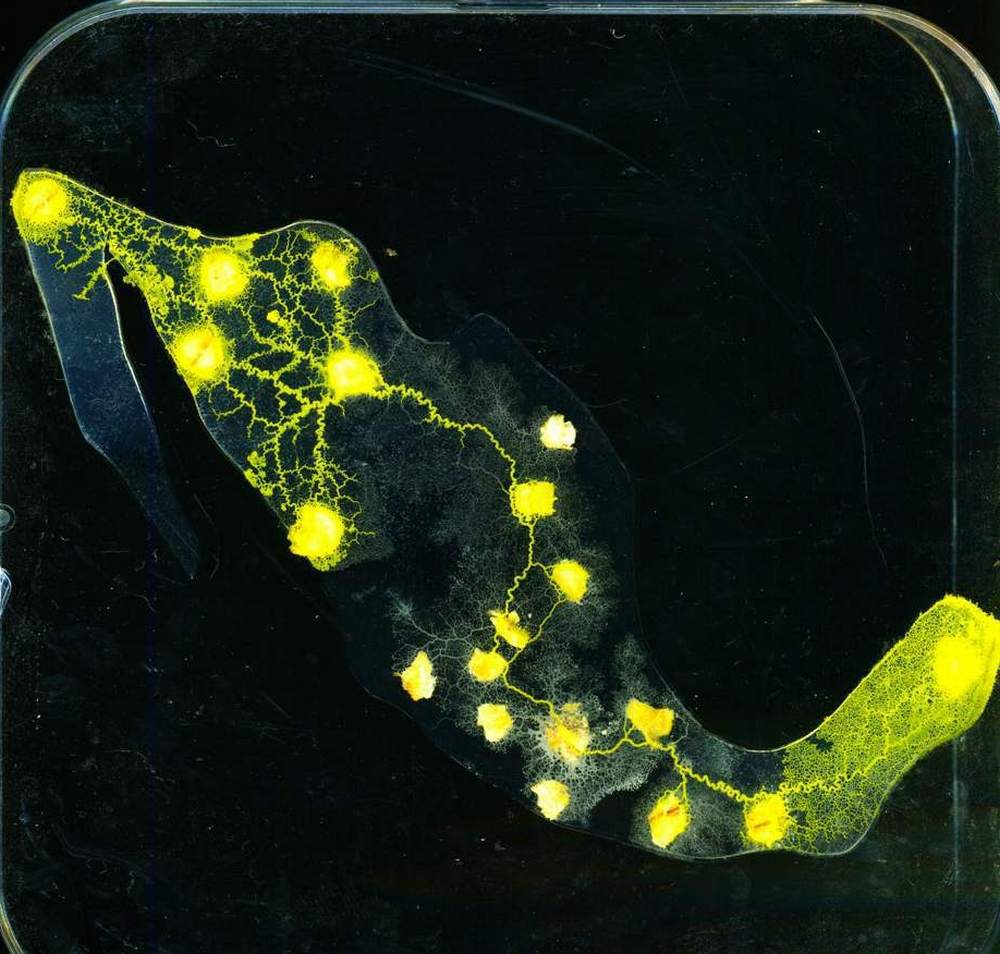}}
\captcont{Plasmodium does not always span all cities (sources of food):
(a)--(d)~scanned image of experimental Petri dish. Time elapsed from inoculation is shown in the sub-figure captions. (e)--(h)~binarized images, $\Theta=(100,100,100)$.}
\label{e4}
\end{figure}

\begin{figure}[!tbp]
\centering
\subfigure[$t=$12~h]{\includegraphics[width=0.4\textwidth]{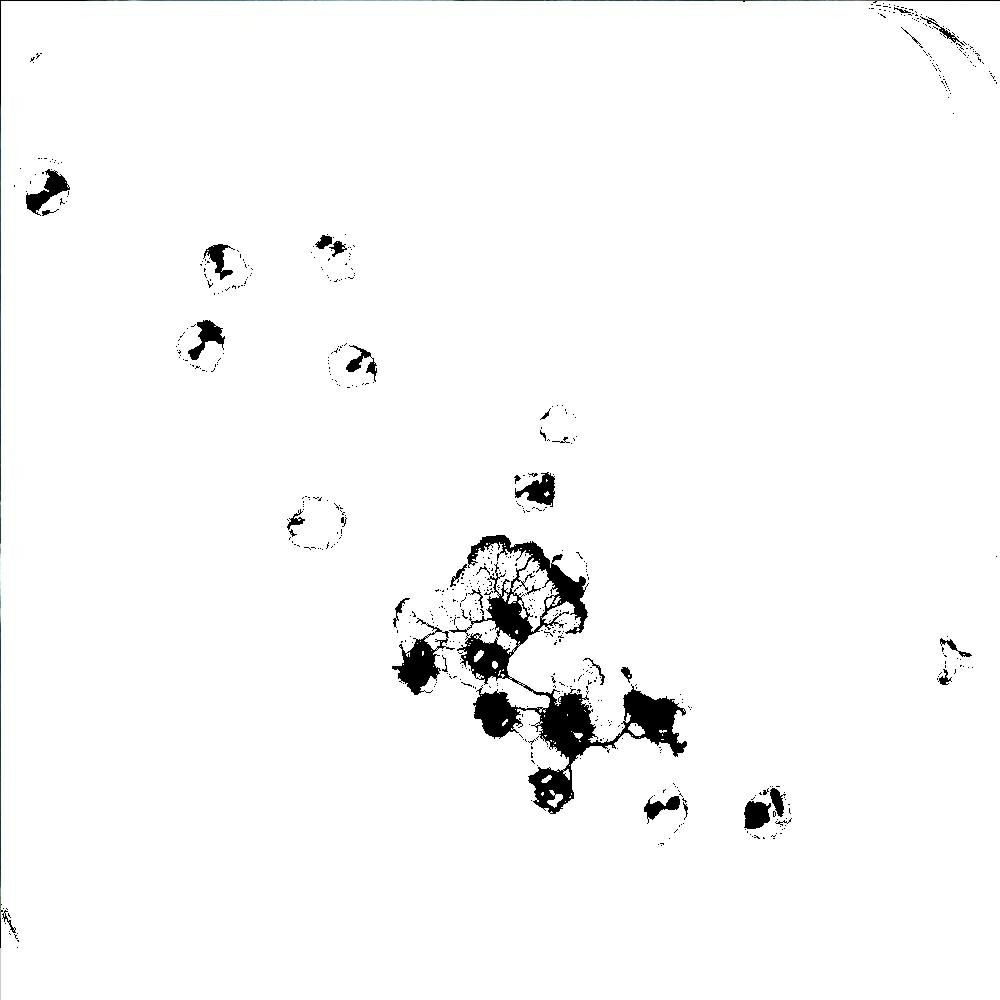}}
\subfigure[$t=$24~h]{\includegraphics[width=0.4\textwidth]{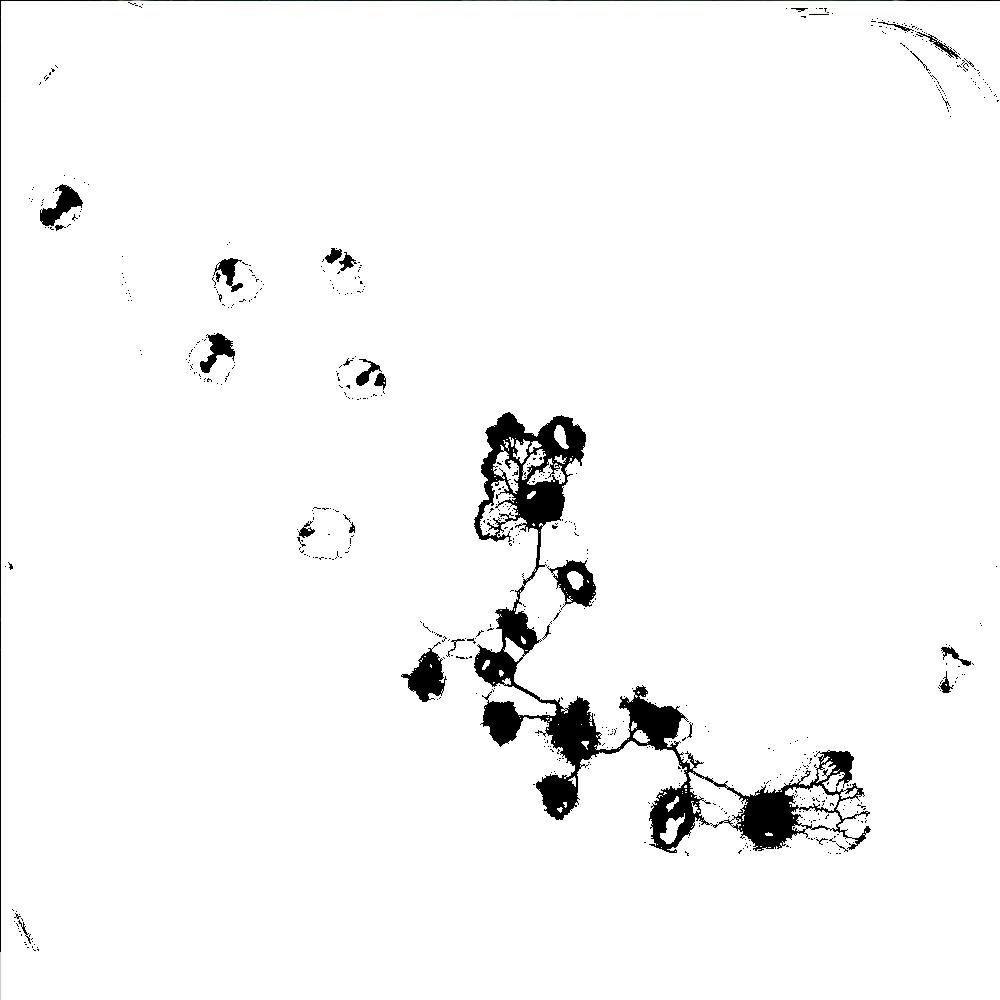}}
\subfigure[$t=$35~h]{\includegraphics[width=0.4\textwidth]{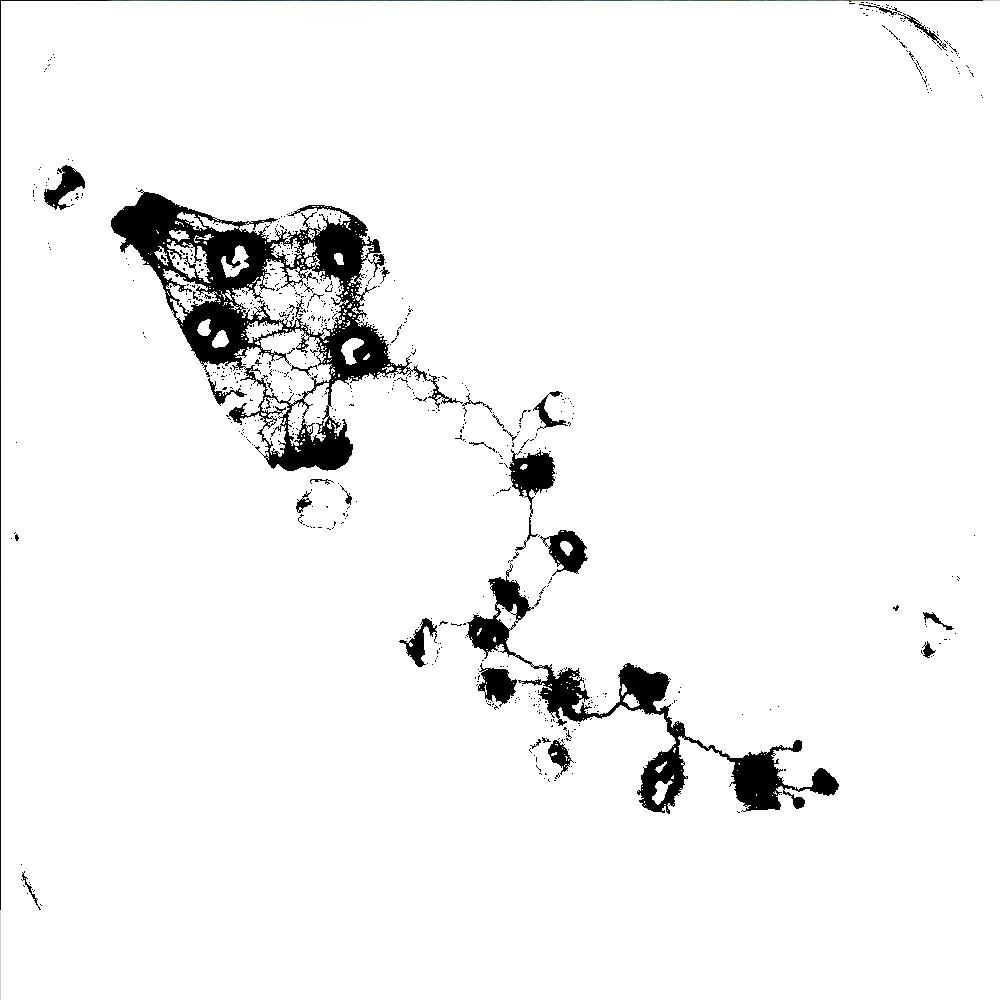}}
\subfigure[$t=$47~h]{\includegraphics[width=0.4\textwidth]{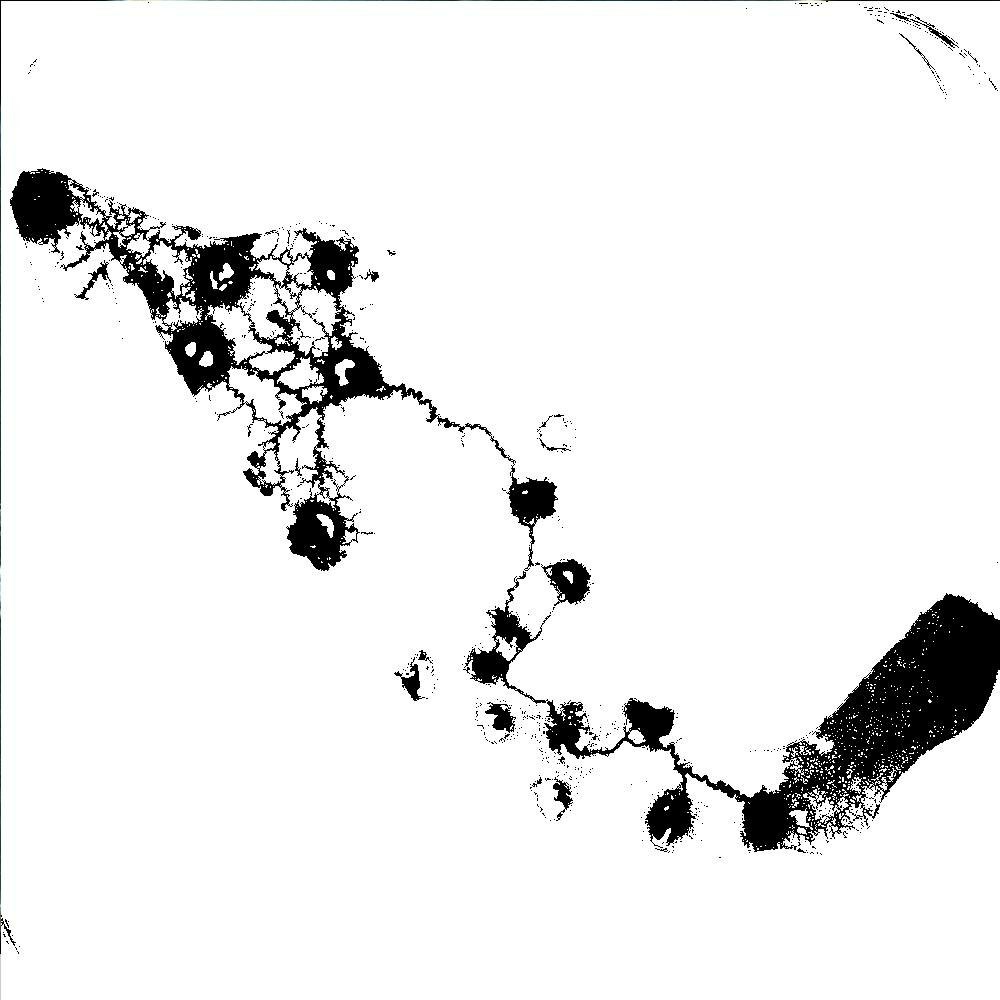}}
\caption{Continued.}
\label{e4}
\end{figure}

In some situations the plasmodium does not span all cities, represented by oat flakes. An example
is shown in (Fig.~\ref{e4}).  In few hours after being inoculated in Mexico city the plasmodium
propagates in all directions. In 12 hours the plasmodium reaches Ciudad Victoria region on the north,
and Xalapa-Veracruz on the south (Fig.~\ref{e4}ae). On 24th hour of development the plasmodium expands
its occupation till Nuevo Laredo region on the north and Tuxtla Guti\'errez on the south (Fig.~\ref{e4}bf). 
The plasmodium continues then to Chihuahua and Merida-Canc\'{u}n regions (Fig.~\ref{e4}cg) and abandons
the oat flake, representing Nuevo Laredo region. Even when the whole territory of Mexico becomes colonized by 
plasmodium the Nuevo Laredo regions remains free (Fig.~\ref{e4}dh). 

\begin{figure}[!tbp]
\centering
\subfigure[$t=$24~h]{\includegraphics[width=0.4\textwidth]{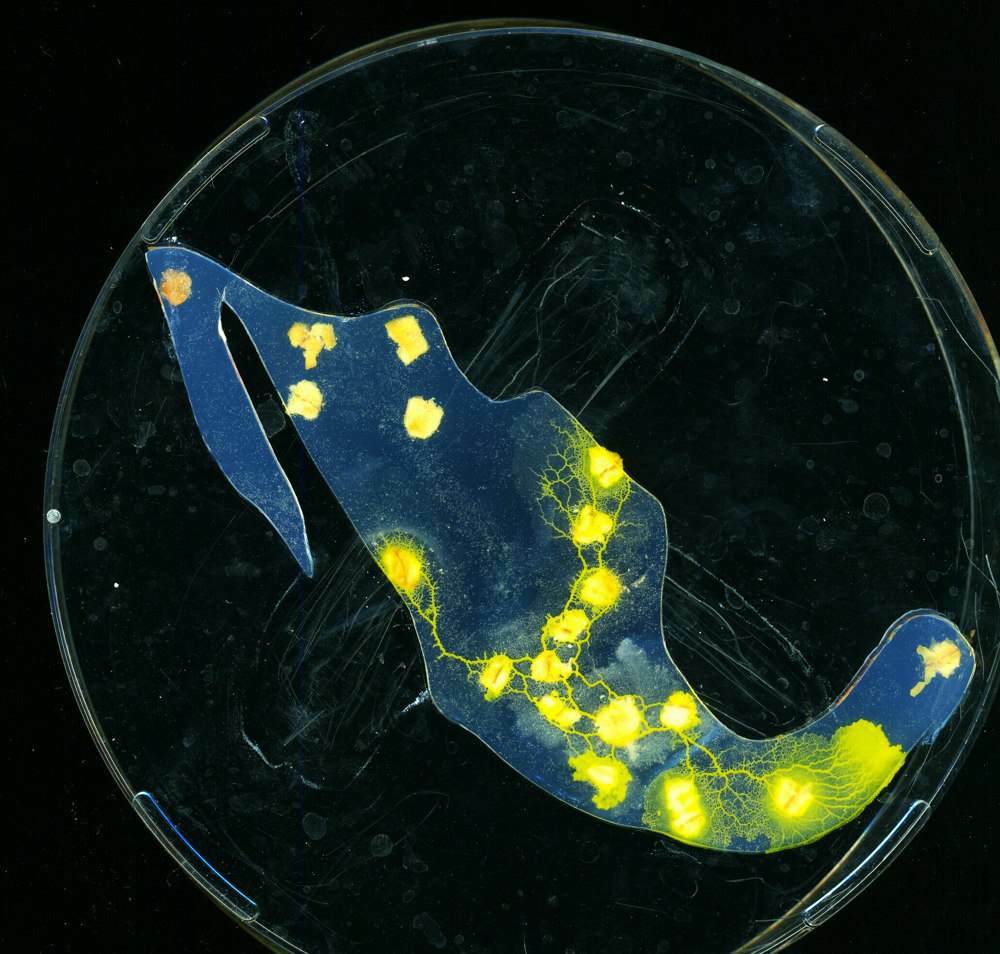}}
\subfigure[$t=$36~h]{\includegraphics[width=0.4\textwidth]{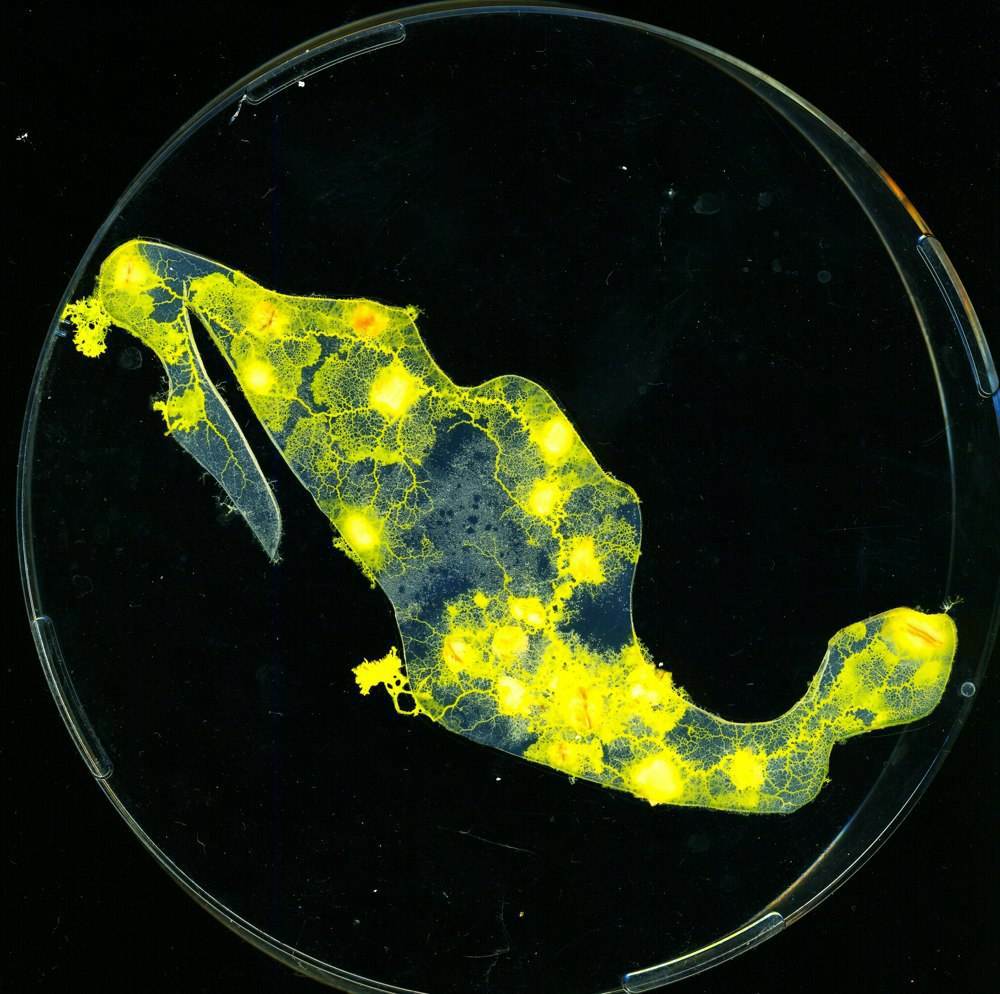}}
\subfigure[$t=$48~h]{\includegraphics[width=0.4\textwidth]{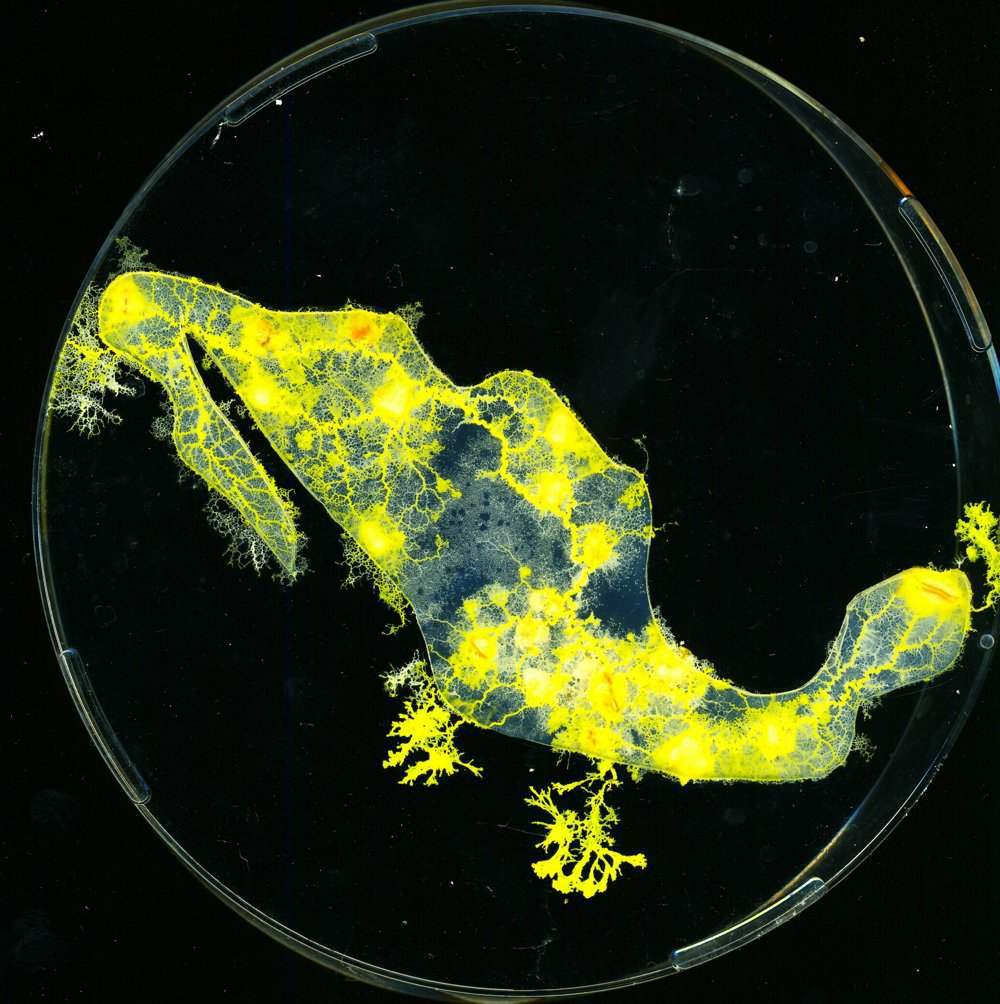}}
\subfigure[$t=$70~h]{\includegraphics[width=0.4\textwidth]{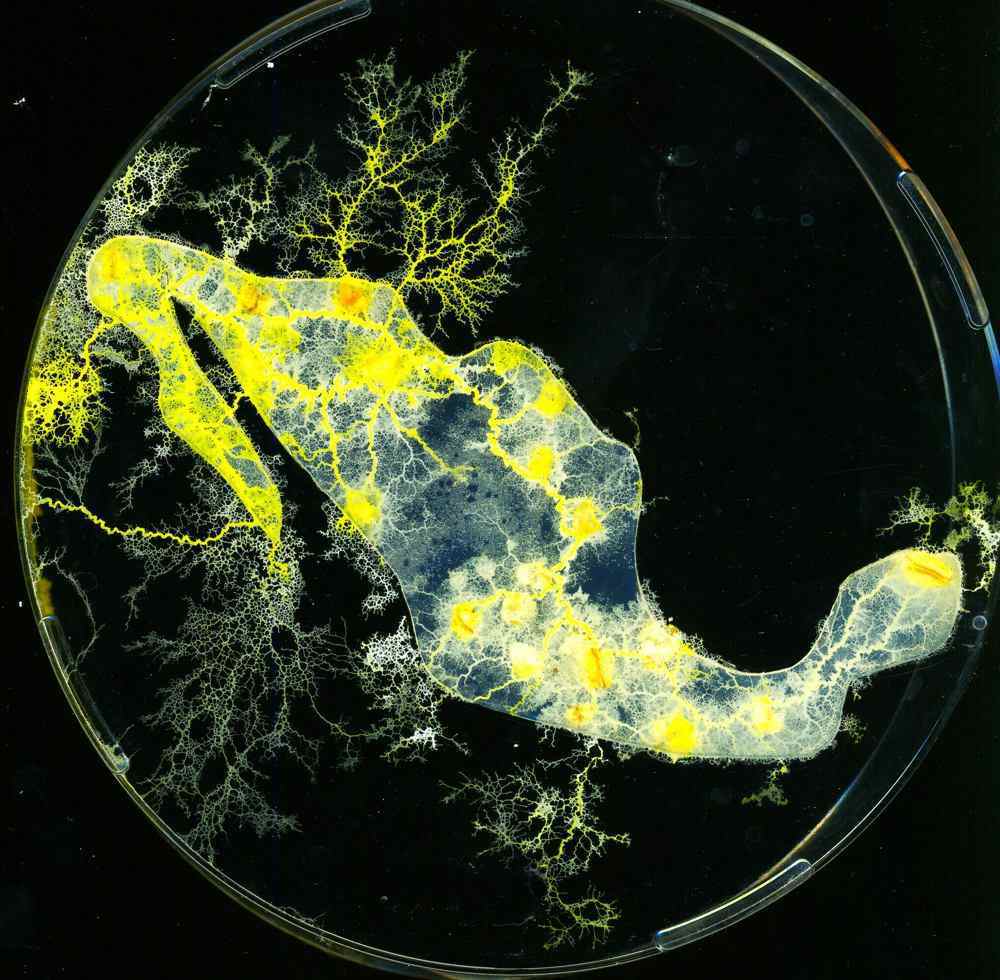}}
\captcont{Plasmodium spreads beyond `dedicated' experimental domain:
(a)--(d)~scanned image of experimental Petri dish. Time elapsed from inoculation is shown in the sub-figure captions. (e)--(h)~binarized images, $\Theta=(100,100,100)$. 
}
\label{m5}
\end{figure}

\begin{figure}[!tbp]
\centering
\subfigure[$t=$24~h]{\includegraphics[width=0.4\textwidth]{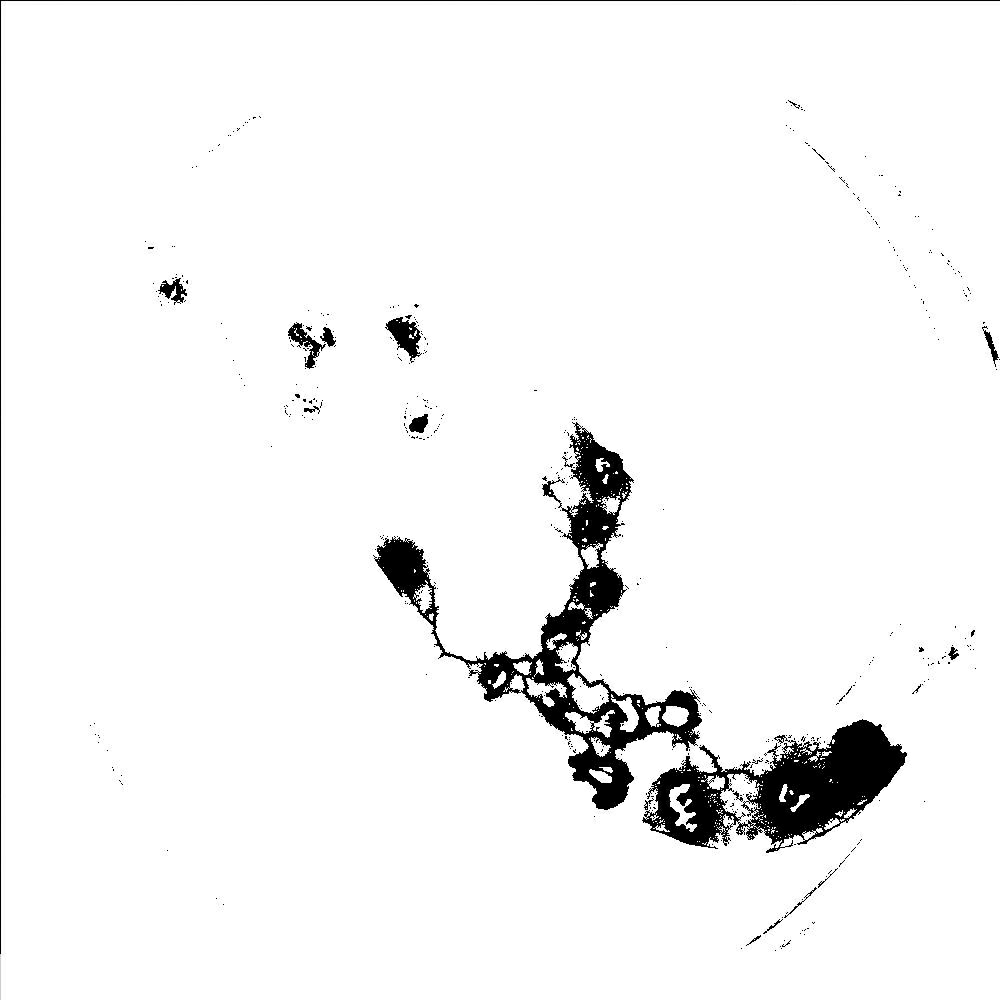}}
\subfigure[$t=$36~h]{\includegraphics[width=0.4\textwidth]{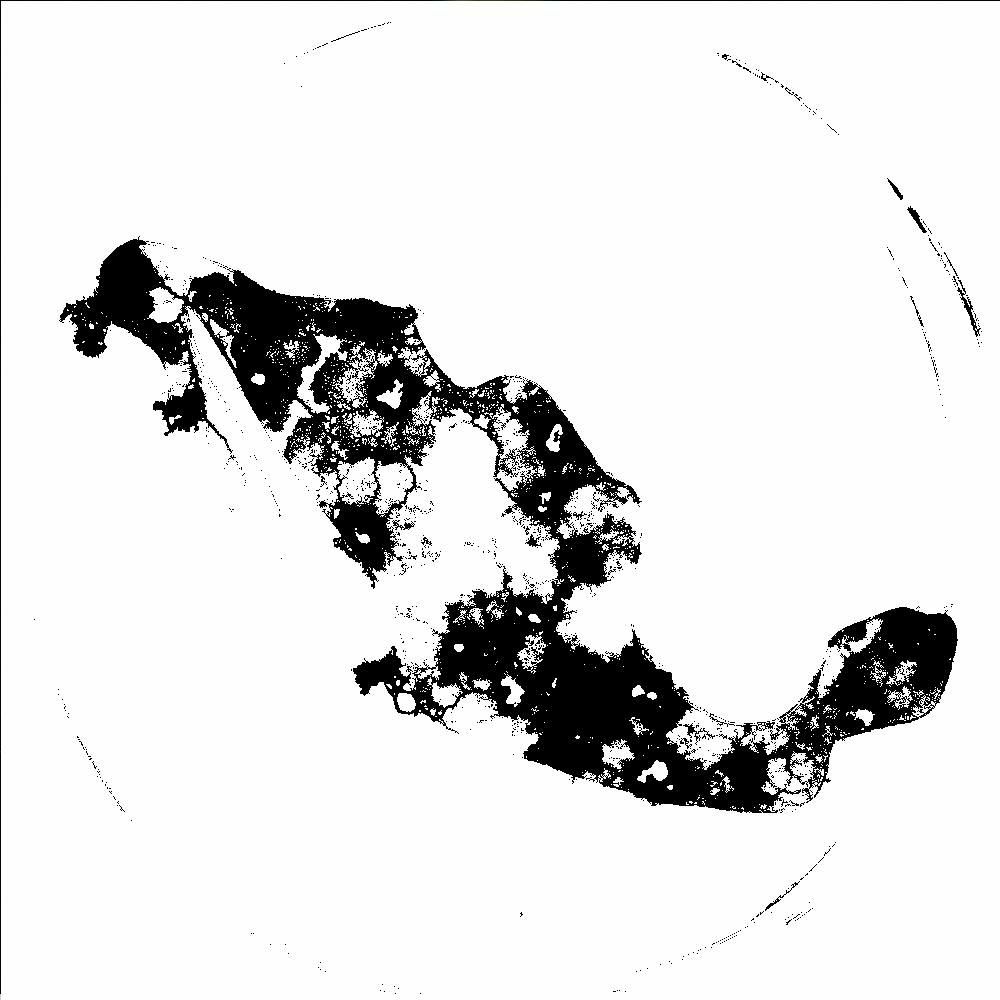}}
\subfigure[$t=$48~h]{\includegraphics[width=0.4\textwidth]{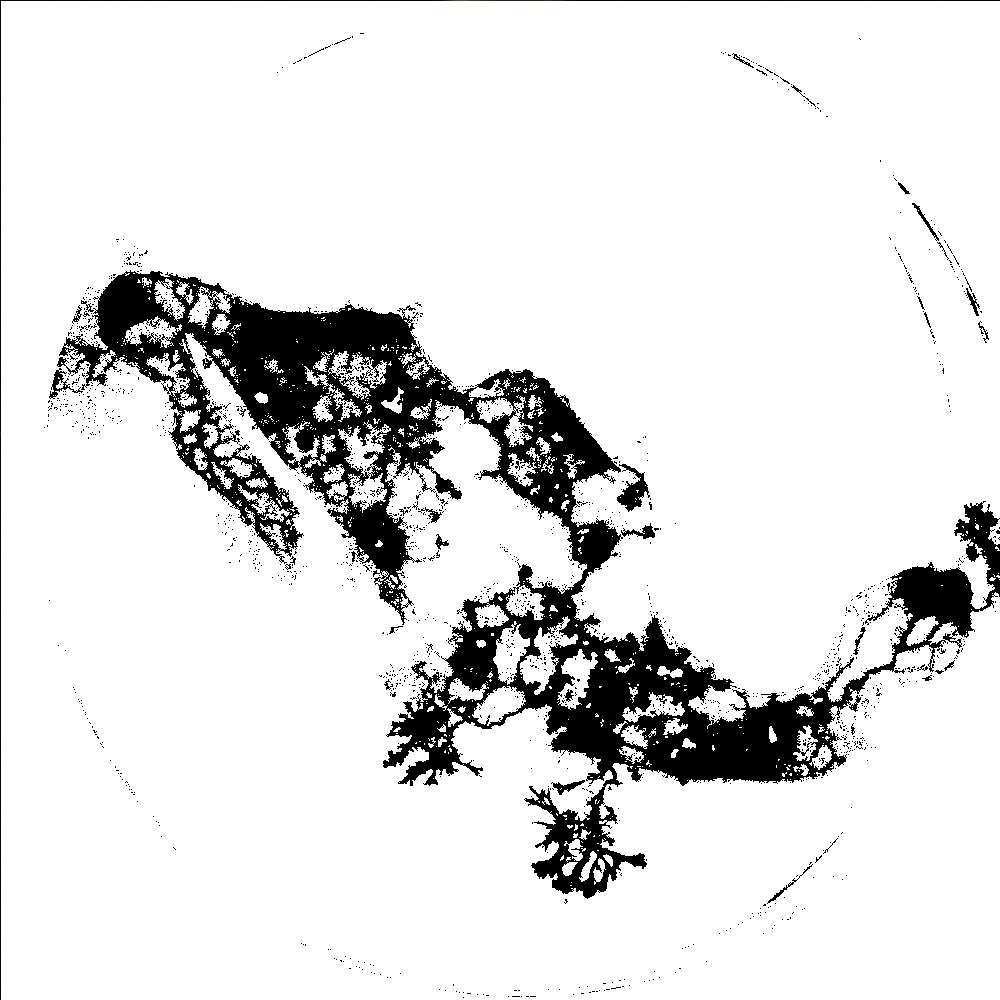}}
\subfigure[$t=$70~h]{\includegraphics[width=0.4\textwidth]{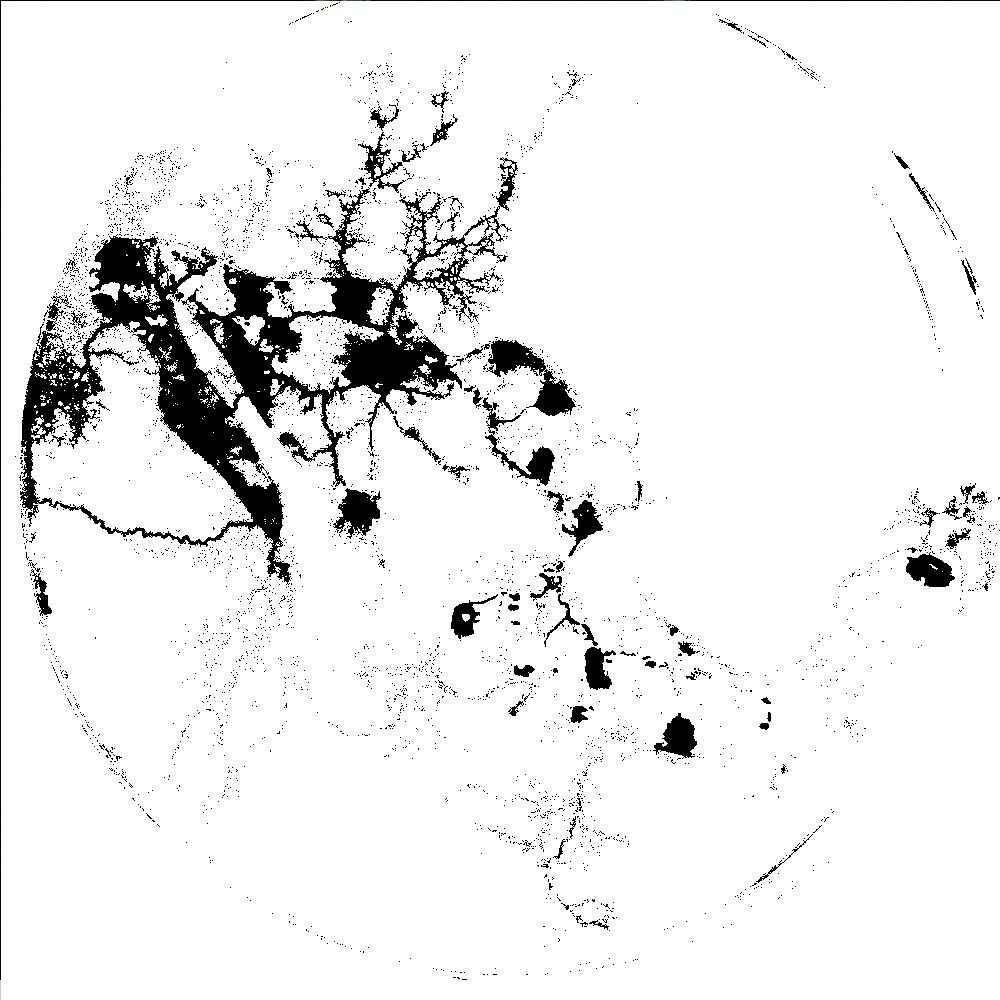}}
\caption{Continued.}
\label{m5}
\end{figure}

The plasmodium may not stop its foraging activity even when all sources of nutrients are occupied and the whole agar plate is explored. As illustrated in Fig.~\ref{m5} a vigorous plasmodium can spread over surrounding Petri dishes, trying to settle on 
a bare plastic.  Plasmodium parts residing on a non-agar substrate suffer from the lack of humidity and therefore cease to 
sustain. Abandoned protoplasmic tubes are clearly visible in southern half of Petri dish in Fig.~\ref{m5}dh. 
  
As illustrated in Figs.~\ref{am9},\ref{e4} and \ref{m5} plasmodium rarely develops exactly the same foraging pattern twice. 
Even in any given experiment the plasmodium sometime may change topology of its protoplasmic networks, abandon and re-colonize sources of food. Thus we could not ever consider a stationary configuration of a protoplasmic network but a probabilistic graph, representing all possible configuration of protoplasmic networks occurring in experiments for a given configuration of 
sources of nutrients. We define a probabilistic Physarum graph in the following manner. 

Physarum graph  is a tuple ${\mathbf P} = \langle {\mathbf U}, {\mathbf E}, w  \rangle$, 
where 
$\mathbf U$ is a set of nineteen urban areas, 
$\mathbf E$ is a set edges, and 
$w: {\mathbf E} \rightarrow [0,1]$ is a probability-weights of edges from $\mathbf E$. 
For every two regions $a$ and $b$ from $\mathbf U$ there is an edge connected $a$ and $b$ if a plasmodium's protoplasmic 
link is recorded at least in one of $k$ experiments, and the edge $(ab)$ has a probability calculated as a ratio of 
experiments where protoplasmic link $(ab)$ occurred to the total number of experiments $k$. We do not take into account exact 
configuration of the protoplasmic tubes but merely their existence. We also consider threshold $\theta \in [0,1]$ Physarum graphs ${\mathbf P}(\theta)$, defined as follows: for $a,b \in \mathbf U$ \, $(ab) \in \mathbf E$ if $w(ab) \geq \theta$.

\begin{figure}[!tbp]
\centering
\subfigure[]{\includegraphics[width=0.4\textwidth]{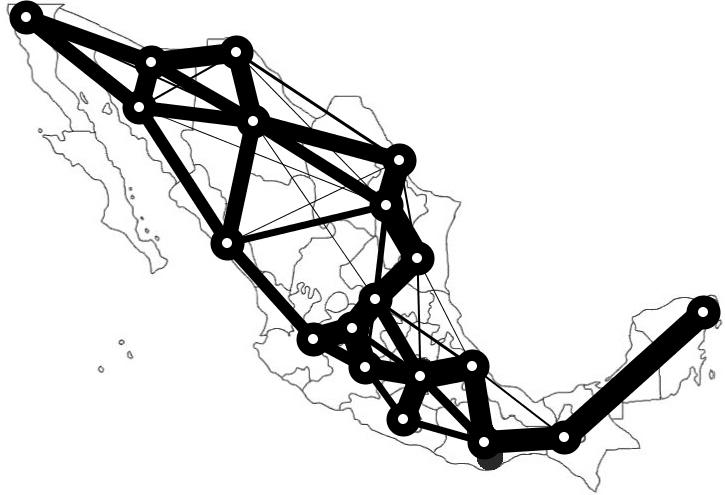}}
\subfigure[]{\includegraphics[width=0.4\textwidth]{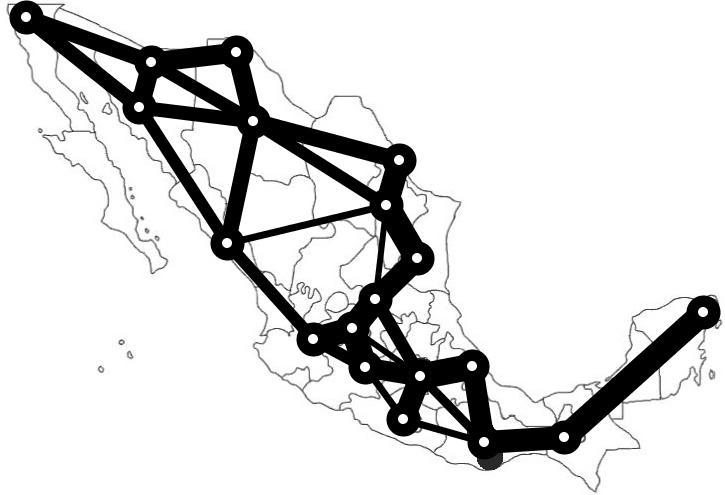}}
\subfigure[]{\includegraphics[width=0.4\textwidth]{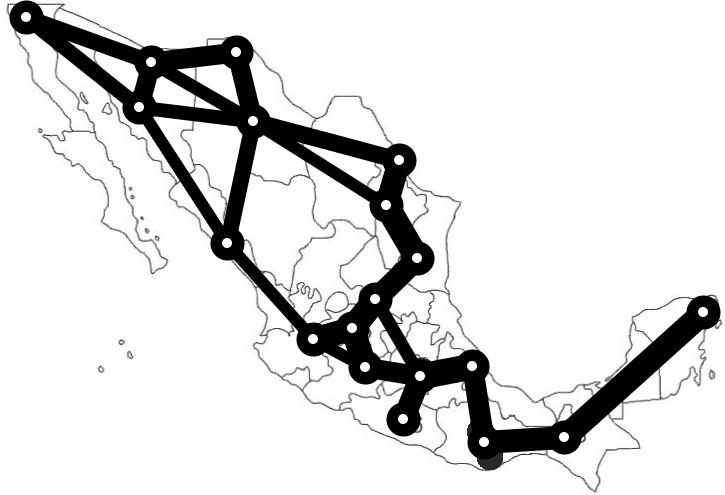}}
\subfigure[]{\includegraphics[width=0.4\textwidth]{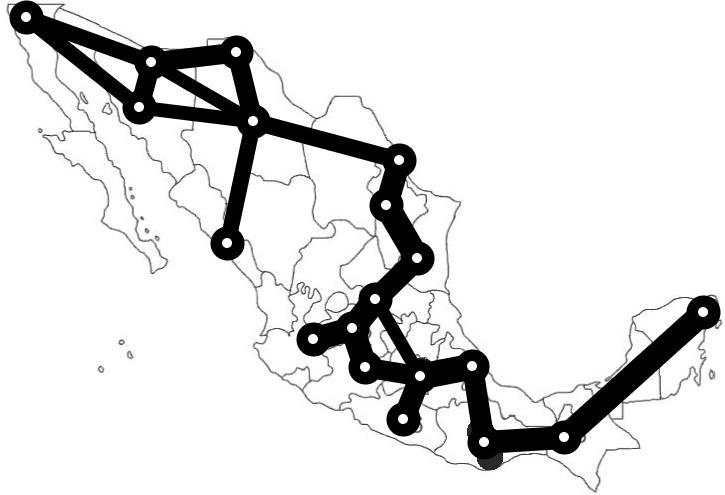}}
\subfigure[]{\includegraphics[width=0.4\textwidth]{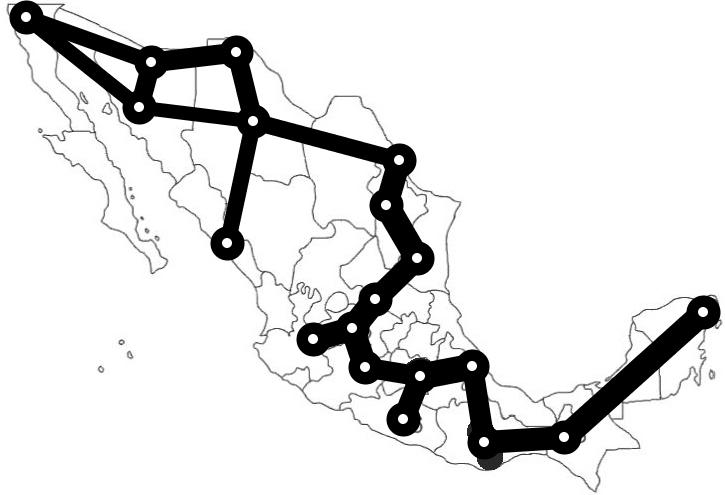}}
\subfigure[]{\includegraphics[width=0.4\textwidth]{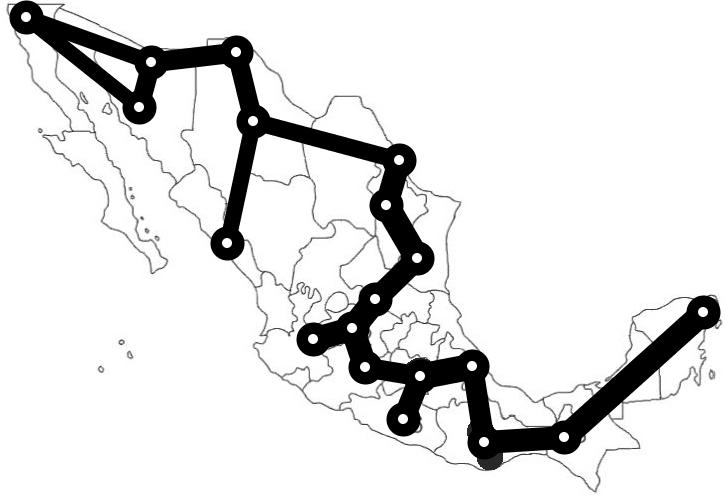}}
\subfigure[]{\includegraphics[width=0.4\textwidth]{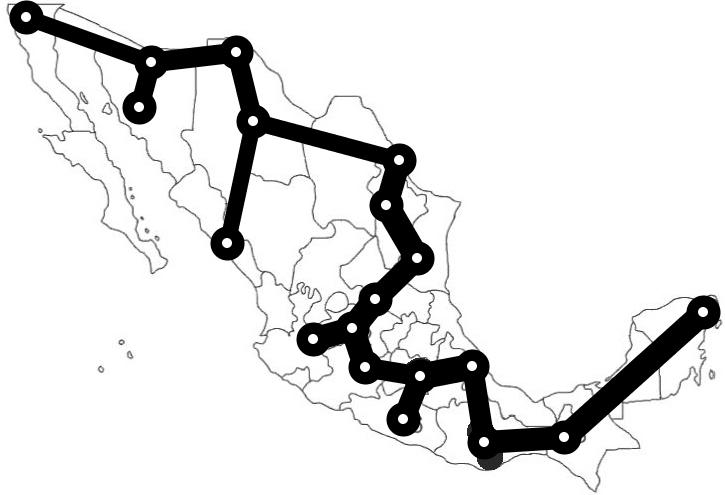}}
\caption{Configurations of threshold \emph{Physarum}-graph ${\mathbf P}(\theta)$ 
for (a)~$\theta=0$,  (b)~$\theta=0.19$, (c)~$\theta=0.38$, (d)~$\theta=0.33$, 
(e)~$\theta=0.5$, (f)~$\theta=0.54$,  (g)~$\theta=0.58$. Thickness of each edge 
is proportional to the edge's weight.}
\label{Fgraphs}
\end{figure}

\begin{figure}[!tbp]
\centering
\subfigure[]{\includegraphics[width=0.4\textwidth]{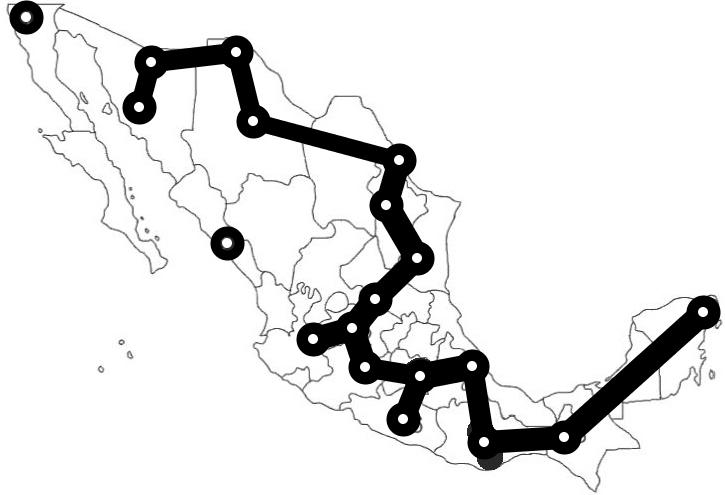}}
\subfigure[]{\includegraphics[width=0.4\textwidth]{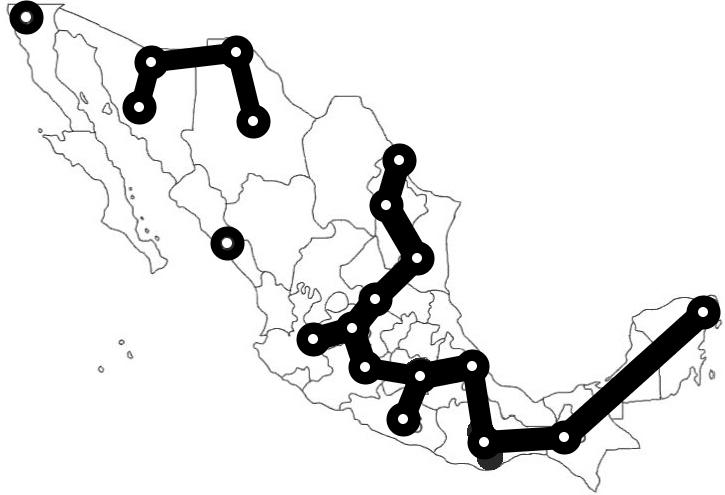}}
\caption{Configurations of threshold \emph{Physarum}-graph ${\mathbf P}(\theta)$ 
for (a)~$\theta=0.61$ and  (b)~$\theta=0.65$. Thickness of each edge 
is proportional to the edge's weight.}
\label{FgraphsDisconnected}
\end{figure}

\emph{Physarum}-graphs extracted from 26 laboratory experiments are shown in Fig.~\ref{Fgraphs}.
The graph becomes planar when we remove edges with weights below 0.19 (Fig.~\ref{Fgraphs}b). 
The graphs is acyclic, or a tree, when only edges with probability-weights exceeding 0.58 are 
taken into consideration (Fig.~\ref{Fgraphs}d). Thus edges of the spanning tree are represented by 
protoplasmic tubes in over half of experimental trials. 

\begin{finding}
Spanning tree is a stable core structure of Physarum foraging and nutrient transport network built 
on a configuration of urban regions $\mathbf  U$.
\end{finding}

If we increase threshold value $\theta$ to 0.61 the Physarum transport graph becomes disconnected (Fig.~\ref{FgraphsDisconnected}a). The nodes corresponding to Tijuana and Mazatl\'{a}n regions becomes isolated. 
There is no a direct highway between such cities, mainly with main big highways as M15 or M2 and this 
isolation of course, represents geographical limitations.

The graph is split into two connected components when $\theta=0.65$ (Fig.~\ref{FgraphsDisconnected}b). Superior component reflects directly the highway dominating north-west of Mexico that is M2 crossing from Tijuana to Ciudad Ju\'arez. The component also represents the fact that Indeed M2 does not reach east coast, that reflect precisely this component. Second component display a strong connection between Mexico city and south-east of Mexico. Such relation involves a high-through  connection region of north-east with Mexico city, and Mexico city with south-east, they are connected specifically by motorway M85 from Nuevo Laredo to Mexico city and M180 that run from Xalapa-Veracruz to Canc\'un.

\begin{figure}[!tbp]
\centering
\includegraphics[width=1.0\textwidth]{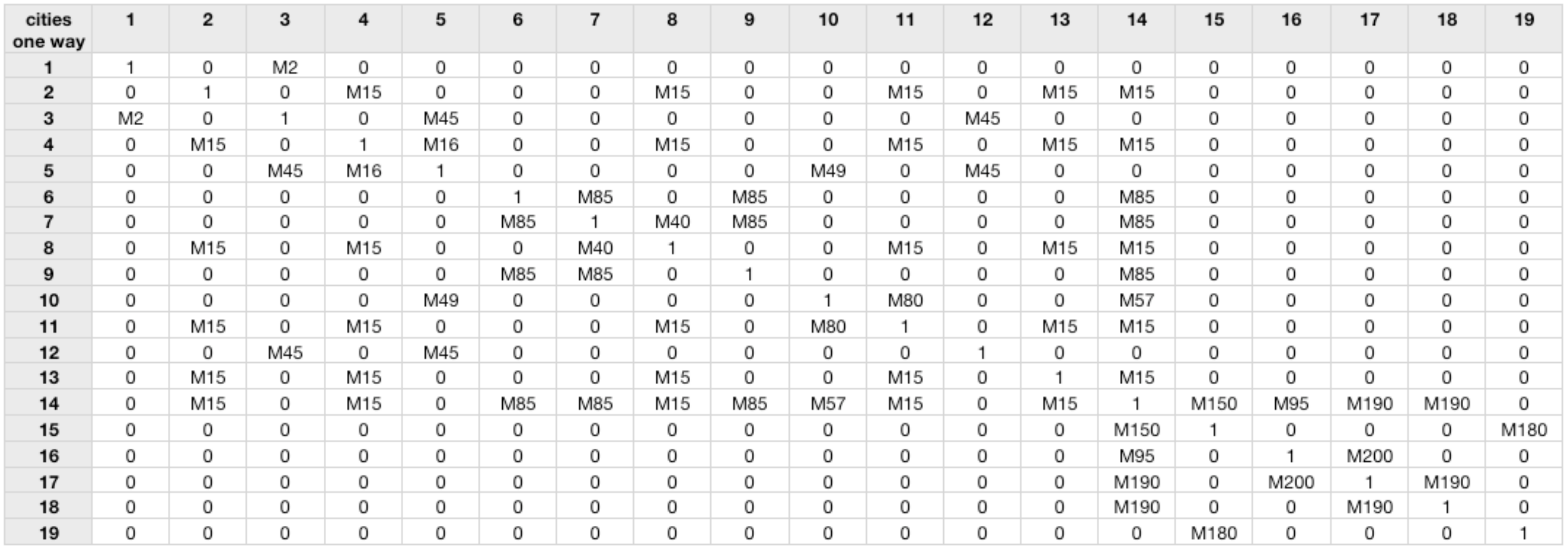}
\caption{Main highways connections between cities of $\mathbf U$.}
\label{motorwaysConnections}
\end{figure}

A specific relation of connections by standard highways (not rural ways) is given in the table~\ref{motorwaysConnections} given from the last update from ``Gu{\'i}a Roji'' publications.\footnote{Source: Gu{\'i}a Roji ``Por las Carreteras de M\'exico 2011'', 17a. Edition. Web site: \url{http://www.guiaroji.com.mx}}

\begin{finding}
Transport links to Tijuana and Mazatl\'{a}n urban regions are unstable. 
\end{finding}

As we saw some implication in Finding 1 (first component), and table in Fig.~\ref{motorwaysConnections} connectivity. We will need at least two or three highways to reach Mazatl\'an from Tijuana. Here there is not a direct motorway as could be M15. Although there is a number of connections to travel between both cities.

\begin{figure}[!tbp]
\centering
\subfigure[]{\includegraphics[width=0.49\textwidth]{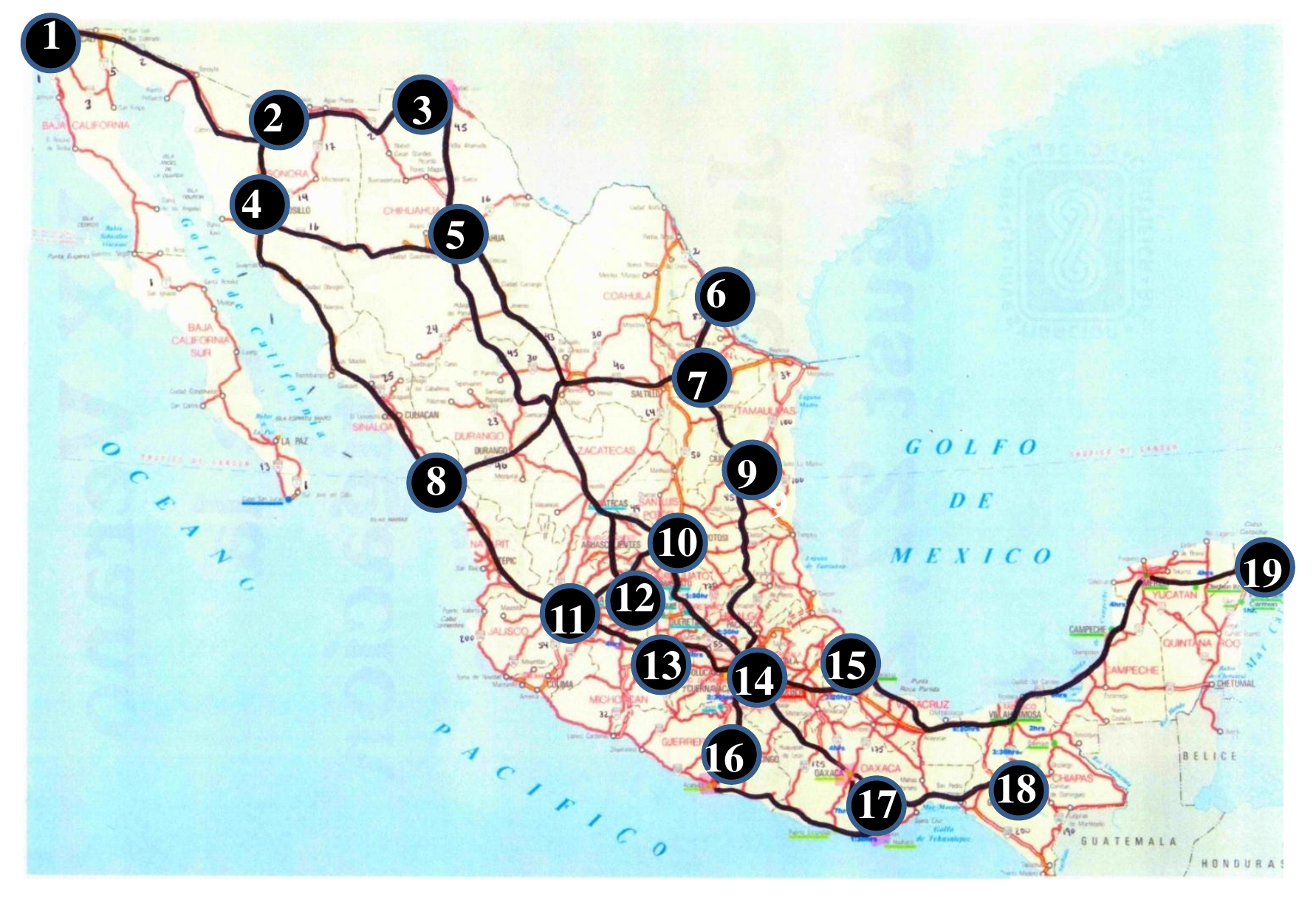}}
\subfigure[]{\includegraphics[width=0.49\textwidth]{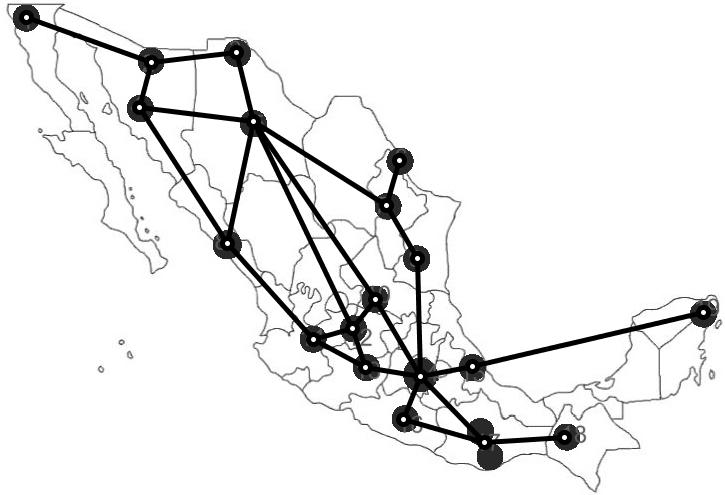}}
\subfigure[]{\includegraphics[width=0.49\textwidth]{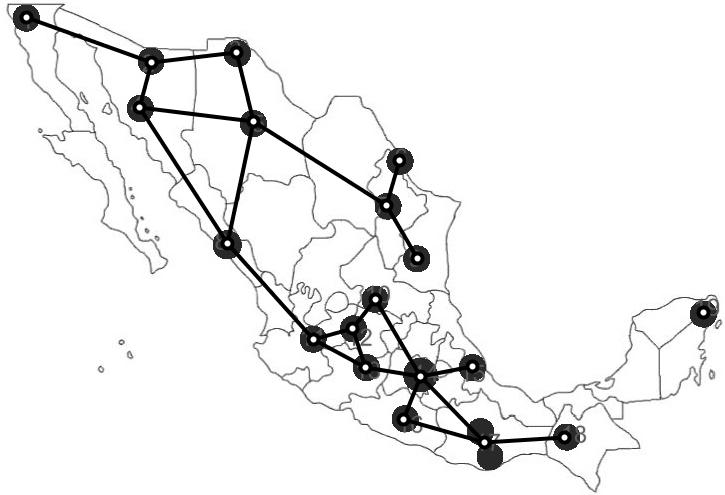}}
\caption{Highway network is highlighted in (a) and graph $\mathbf H$ of man-made motorway network is shown in (b). Intersection ${\mathbf P} \bigcap {\mathbf H}$ of Physarum $\mathbf P$ and highway $\mathbf H$ graphs is shown in (c).}
\label{motorways}
\end{figure}

On the way, how well Physarum graphs approximate Mexican highway network. A sketch of the highway network and the graph $\mathbf H$ derived are shown in Fig.~\ref{motorways}. We construct the highway graph $\mathbf H$ as follows. Let $\mathbf U$ be a set of nineteen urban regions, for any two regions $a$ and $b$ from $\mathbf U$, the nodes $a$ and $b$ are connected by an edge $(ab)$ if there is a highway starting in vicinity of $a$ and passing in vicinity of $b$ and not passing in vicinity of any other urban area $c \in \mathbf U$.   

\begin{finding}
Graph ${\mathbf P}(0.58)$ is a sub-graph of $\mathbf H$ apart of three edges.
\end{finding}

Intersection ${\mathbf P} \bigcap {\mathbf H}$ of Physarum and highway graphs is shown in Fig.~\ref{motorways}c. The three edges of ${\mathbf P}(0.58)$ missing in  $\mathbf H$ are transport links connecting Chihuahua and Nuevo Laredo regions, Xalapa-Veracruz and Oaxaca-Huatulco regions, and Tuxtla Guti\'errez and Merida-Canc\'{u}n regions. Such difference could possibly be explained by the fact that in laboratory experiments we did not represent geographical landscape in agar gel structure but just a uniform gel plate was cut in a shape of Mexico.   

Indeed, we have considered natural restrictions. Missing connections between Chihuahua and Nuevo Laredo are limited by Chihuahua desert. While connection between Xalapa-Veracruz and Oaxaca-Huatulco mainly is limited by geographical features such as Ju\'arez mountain range. So missing connection between Tuxtla Guti\'errez and Merida-Canc\'un implies a large border over Golf of Mexico upping to motorway M180 and crossing also on Chiapas mountain range.

\section{Physarum, proximity graphs and highways}
\label{proximitygraphs}

\begin{figure}[!tbp]
\centering
\subfigure[$\mathbf{RNG}$]{\includegraphics[width=0.4\textwidth]{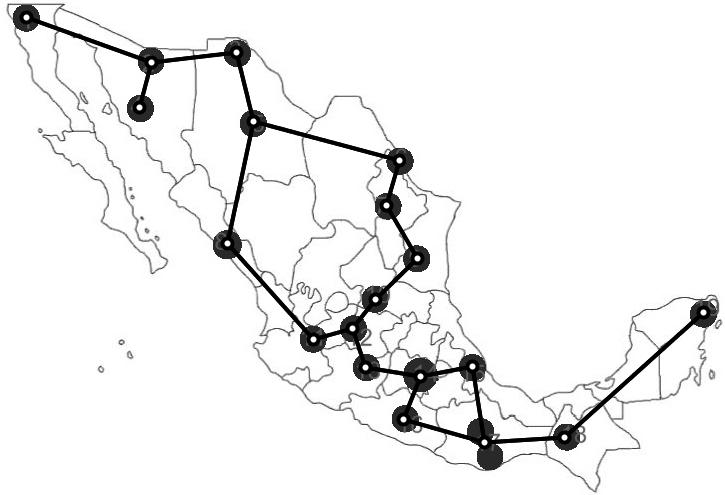}}
\subfigure[$\mathbf{BS}$(1.5)]{\includegraphics[width=0.4\textwidth]{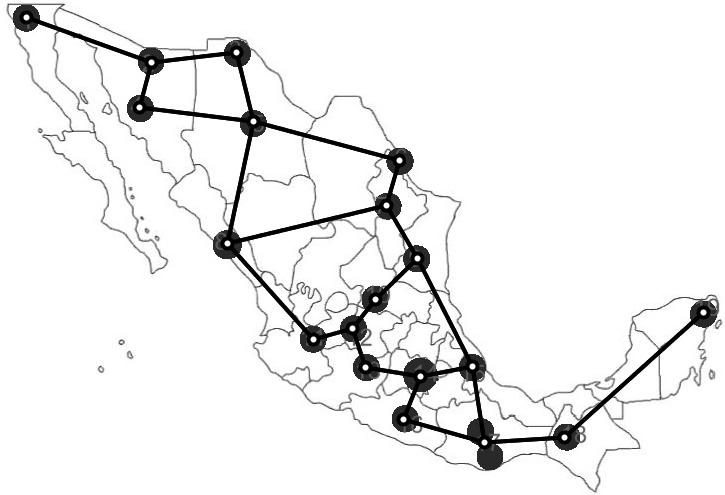}}
\subfigure[$\mathbf{GG}$]{\includegraphics[width=0.4\textwidth]{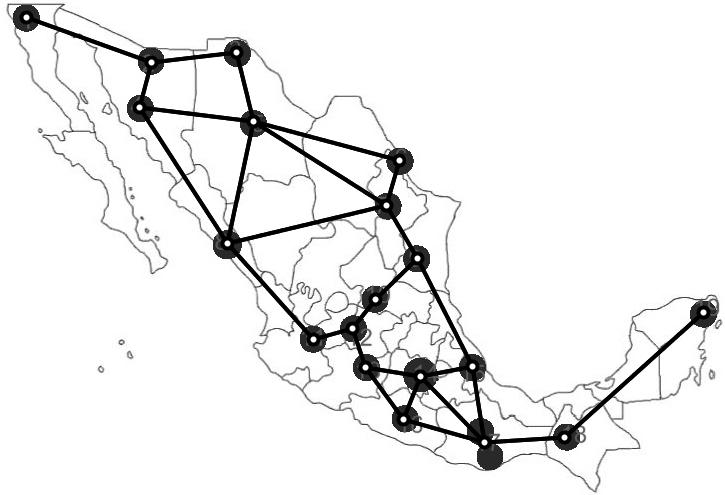}}
\subfigure[$\mathbf{ST}$]{\includegraphics[width=0.4\textwidth]{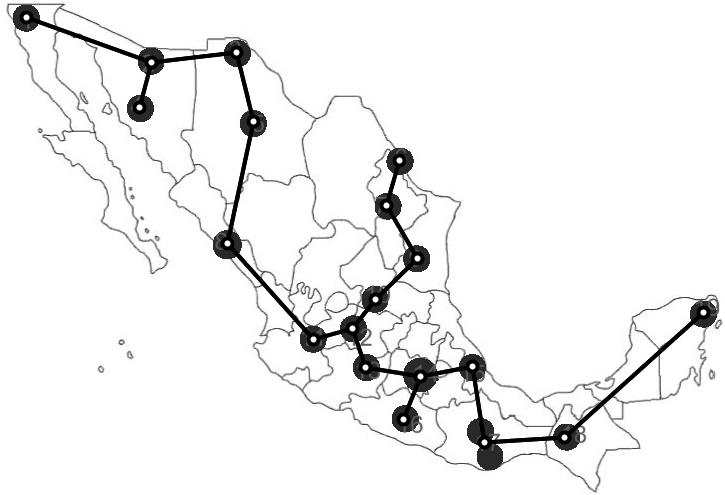}}
\caption{Proximity graphs constructed on regions $\mathbf U$: 
(a)~relative neighborhood graph $\mathbf{RNG}$,
(b)~$\beta$-skeleton with control parameter 1.5,
(c)~Gabriel graph $\mathbf{GG}$,
(d)~Minimum spanning tree $\mathbf{ST}$}
\label{proximity}
\end{figure}

Physarum constructs planar proximity graphs by its protoplasmic network~\cite{adamatzky_ppl_2009}. A planar graph consists of nodes which are points of Euclidean plane and edges which are straight segments connecting the points. A planar proximity graph is a planar graph where two points are connected by an edge if they are close in some sense. Usually a pair of points is assigned certain neighborhood, and points of the pair are connected by an edge if their neighborhood is empty.  
Relative neighborhood graph~\cite{jaromczyk_toussaint_1992}, Gabriel graph~\cite{matula_sokal_1984}, $\beta$-skeletons~\cite{kirkpatrick} 
and spanning tree are most known examples of proximity graphs. 

Relative neighborhood graph~\cite{toussaint_1980} $\mathbf{RNG}$ (Fig.~\ref{proximity}a) and Gabriel graph~\cite{gabriel_sokal_1969, matula_sokal_1984} $\mathbf{GG}$ (Fig.~\ref{proximity}b) are computed 
over nodes corresponding to  urban areas $\mathbf U$. Points $a$ and $b$ are connected by an edge 
in $\mathbf{RNG}$ if no other point $c$ is closer to $a$ and $b$ than $dist(a,b)$~\cite{toussaint_1980}. 
Points $a$ and $b$ are connected by edge in $\mathbf{GG}$ if a disc with diameter $dist(a,b)$ centered in middle of the segment $ab$ is empty~\cite{gabriel_sokal_1969, matula_sokal_1984}.
The graphs are related as  $\mathbf{RNG} \subseteq \mathbf{GG}$~\cite{toussaint_1980,matula_sokal_1984,jaromczyk_toussaint_1992}.

\begin{figure}[!tbp]
\centering
\subfigure[$\mathbf P \bigcap \mathbf{ RNG}$ ]{\includegraphics[width=0.45\textwidth]{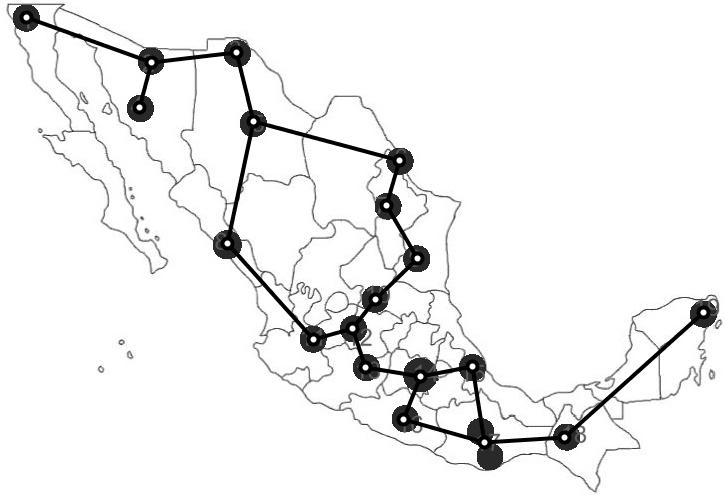}}
\subfigure[$\mathbf P \bigcap \mathbf{ BS}$(1.5)]{\includegraphics[width=0.45\textwidth]{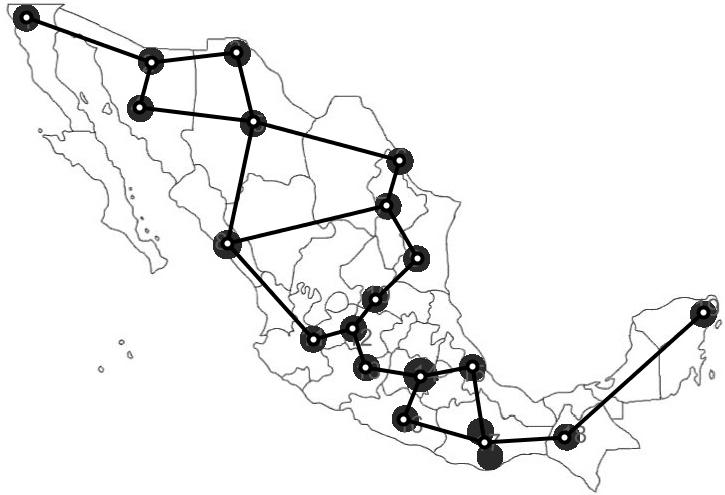}}
\subfigure[$\mathbf P \bigcap \mathbf{ GG}$]{\includegraphics[width=0.45\textwidth]{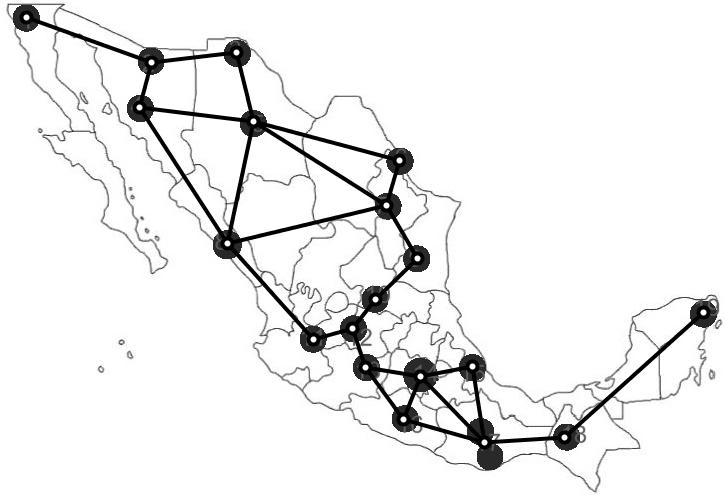}}
\subfigure[$\mathbf P \bigcap \mathbf{ MST}$]{\includegraphics[width=0.45\textwidth]{figs/202553_ST}}
\subfigure[$\mathbf H \bigcap \mathbf{ RNG}$ ]{\includegraphics[width=0.45\textwidth]{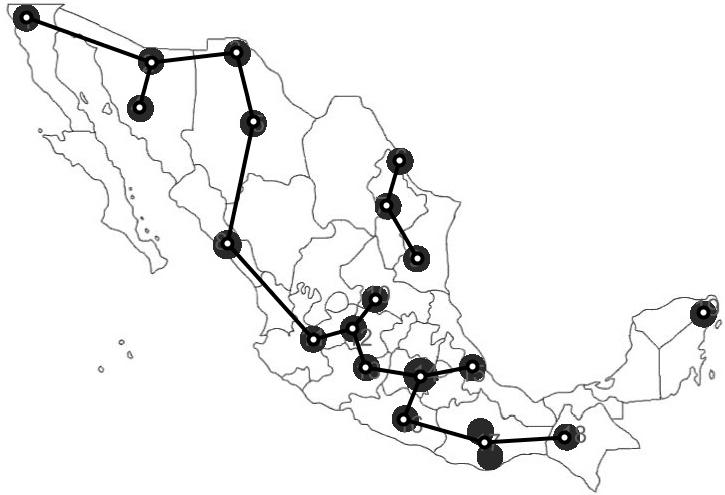}}
\subfigure[$\mathbf H \bigcap \mathbf{ BS}$(1.5)]{\includegraphics[width=0.45\textwidth]{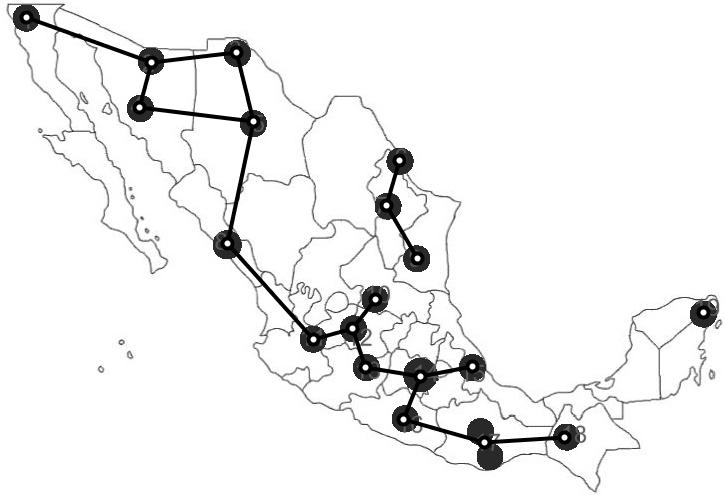}}
\subfigure[$\mathbf H \bigcap \mathbf{ GG}$]{\includegraphics[width=0.45\textwidth]{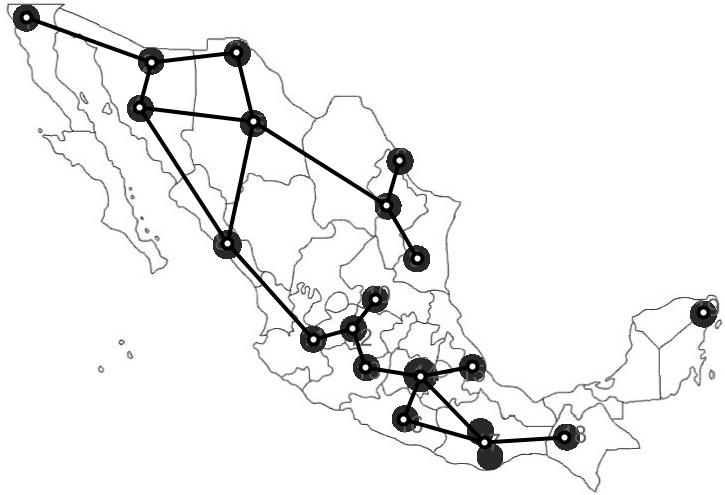}}
\subfigure[$\mathbf H \bigcap \mathbf{ MST}$]{\includegraphics[width=0.45\textwidth]{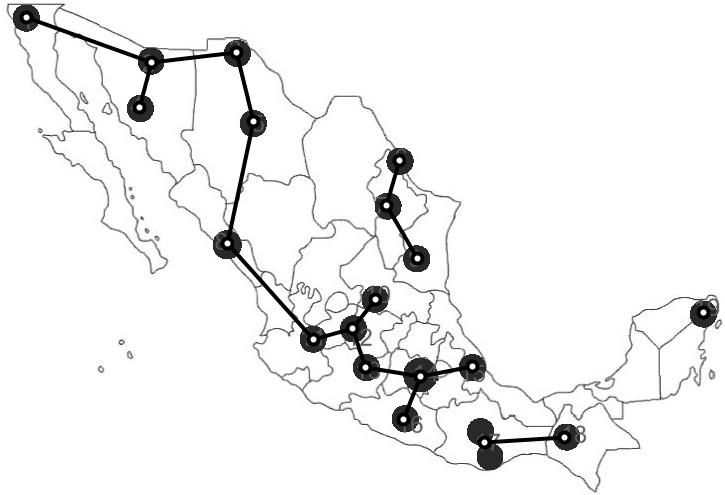}}
\caption{Intersection of Physarum graph $\mathbf P$~(a)--(d) and highway graph $\mathbf H$~(e)--(h) with proximity relative neighborhood graph $\mathbf{ RNG}$, $\beta$-skeleton $\mathbf{ RNG}$(1.5), Gabriel graph $\mathbf{ GG}$ and minimum spanning tree $\mathbf{ MST}$.}
\label{intersections}
\end{figure}

Proximity graphs found their applications in many fields of science and engineering. The application domains related
to topics of the paper include routing in ad hoc wireless networks~\cite{li_2004,song_2004,santi_2005,muhammad_2007,wan_2007},
road network analysis~\cite{watanabe_2005, watanabe_2008}, study percolation~\cite{billiot_2010} and analysis of magnetic field~\cite{sridharan_2010}. Structure of proximity graphs represents so wide range of natural and engineering system that 
it would be productive to compare Physarum graph $\mathbf P$ and highway graph $\mathbf H$ to relative neighborhood graph $\mathbf{RNG}$, $\beta$-skeleton $\mathbf{BS}$(1.5) ($\beta$-skeleton $\mathbf B$ was calculated for $\beta=1.5$ to make
an `intermediate' graph between $\mathbf{RNG}$ and $\mathbf{GG}$), Gabriel graph $\mathbf{GG}$ and 
minimum spanning tree $\mathbf{MST}$ (Fig.~\ref{intersections}). From now on we mean $\mathbf{P}(0.19)$ when 
writing $\mathbf P$.

\begin{finding}
$\mathbf{RNG} \subset \mathbf{P}$
\end{finding}

This is a direct outcome of calculating intersections of the graphs (Fig.~\ref{intersections}). Relative neighborhood graph is considered to be optimal in terms of total edge length and travel distance, and is a closest approximation of road networks~\cite{watanabe_2005, watanabe_2008}. The fact this graph is a sub-graph of Physarum graph indicates deep intrinsic analogies between formation of man-made and plasmodium transport networks.

In all cases (Fig.~\ref{intersections}a--d) the large pathway between Mexico city and USA is preserved in three following ways. Mexico city via Ciudad Ju\'arez by motorway M45, Mexico city via Nogales by motorway M15, and Mexico city via Nuevo Laredo by motorway M85. They are very important since Mexico colonization because they represent ancient ways to transport materials and nature sources to Spanish monarchs and later British colonies (now USA). Of course, they were besides guided initially from Xalapa-Veracruz to Mexico city (from Aztec's age where Spanish tried to reach Tenochtitlan), and consequently defining this way from old times, that actually represent motorway M150.

In this context it is not surprising that ``Ciudad Ju\'arez'' play a strategic position in reaching USA directly from Mexico center and beyond. This is because motorway M45 is connected with M190 and runs across all Mexican Republic south-to-north from Central America. It is also important to highlight that Ciudad Ju\'arez undergone an explosive growth in last few years, partly to its dramatic life at the intersection of main drug transport arteries. 

South-east intersections represent connectivity with highways M180 and M190. Both are the main large highways 
connecting principal cities in their regions with Mexico city.

\begin{finding}
$\mathbf{MST} \subset \mathbf{P}$
\end{finding}

The result is expected because spanning tree is a sub-graph of a relative neighborhood graph~\cite{toussaint_1980}, however we intentionally highlighted it to show that --- in terms of minimal-length travel --- graph, built by plasmodium of {\emph Physarum polycephalum} offers optimal solution for transportation of nutrients. Experiment-wise we found that the spanning tree $\mathbf{MST}$ is a sub-graph of Physarum graphs $\mathbf{P}(\theta)$ for $\theta \leq 0.58$, compare Fig.~\ref{Fgraphs} and \ref{intersections}d. 

As we demonstrated previously (Finding 1) by cutting edges of $\mathbf{P}(\theta)$ with $\theta \leq 0.58$ we transform cyclic graph to a spanning tree (Fig.~\ref{Fgraphs}g). By comparing Fig.~\ref{Fgraphs}g and Fig.~\ref{proximity}d we find that if Physarum built a transport link connecting Mazatl\'{a}n and Guadalajara regions instead of connecting Chihuahua and Nuevo Laredo regions then the Physarum would approximate an ideal minimum spanning tree. Actual travel distance makes Physarum spanning tree $\mathbf{P}(0.58)$ just a bit longer than ideal spanning tree $\mathbf{MST}$, constructed by conventional algorithm Fig.~\ref{proximity}d.
  
In contrast to Physarum graphs highway graph $\mathbf{H}$ poorly matches proximity graphs (Fig.~\ref{intersections}e--h).

\begin{finding}
Let $\mathbf{G} \in \{\mathbf{RNG}, \mathbf{BS}(1.5), \mathbf{GG}, \mathbf{MST}  \}$ then $\mathbf{G} \bigcap \mathbf{H}$ is a disconnected graph.
\end{finding} 

Merida-Canc\'{u}n region stays isolated in all four intersections (Fig.~\ref{intersections}e--h) of the highway graph $\mathbf{H}$ with proximity graphs. A connected cluster of regions {Nuevo Laredo, Monterrey, Ciudad Victoria} is disconnected from the rest of other vertices of $\mathbf{U}$ in $\mathbf{H} \bigcap \mathbf{RNG}$, $\mathbf{H} \bigcap \mathbf{BS}(1.5)$, and $\mathbf{H} \bigcap \mathbf{GG}$. Of course, this cluster represents the most important economic activity of the region. But in this case the connectivity result is poor with respect to intersection with $\mathbf{P}$. A similar situation takes place with 
Merida-Canc\'un connection in intersection of minimum spanning tree $\mathbf{MST}$ with highway graph $\mathbf{H}$.

Let $a$ be a transport link connecting San Luis Potos{\'i} region with Mexico city-Puebla-State of M\'exico region then we see that links of motorway network $\mathbf H$ which are also edges of Gabriel graph $\mathbf{GG}$ correspond to the common links of the motorway network $\mathbf H$ and Physarum graph $\mathbf{P}(0.19)$. San Luis Potos{\'i} connection represents a special case linked directly to Mexico city region with motorway M57.

\begin{finding}
$\mathbf{H} \bigcap \mathbf{GG} \bigcup \{ a \} = \mathbf{H} \bigcap \mathbf{P}(0.19)$ 
\end{finding} 

Unfortunately, we have lost a link between Merida-Canc\'un and Mexico center and also a center-to-north link 
from San Luis Potos{\'i} to Chihuahua. This means that motorways M85 and M190 are partly misrepresented. 
Said that, the pacific-long highway connecting Nogales to Mexico city by M15 is preserved. Thus, at least, we 
have preserved one of the largest and most important connection between Mexico city and the North of Mexico. 
This way, $\mathbf{P}(0.58)$ is the best approximation as we saw previously.

\section{Discussion}
\label{discussion}

To approximate, or rather re-construct, development of transport network in Mexico we cut of Mexico-shaped plate of agar, represented nineteen major urban regions by oat flakes and placed a plasmodium of {\emph Physarum polycephalum} in place of Mexico city.\footnote{You can see an approximation of these simulations from \url{http://www.youtube.com/watch?v=OmwtPR2cV-4}.} The plasmodium developed into a full-fledged plasmodium spanning all, or almost all, oat flakes on the agar plate. We compared graphs of protoplasmic networks developed by Physarum with man-made Federal Highway network and with basic types of proximity graphs. 

We found  that Physarum-made graph is a sub-graph of the highway network apart of few ages. This means that, in principle, slime mould based approximation of man-made transport networks works even in very simple experimental setups. The slight mismatch between Physarum and highway graphs may be because we did not take geographical profile of Mexico into account completely, no mountains or rivers were mapped onto the plasmodium growth substrate.  Also, inoculation of plasmodium in Mexico city was backed up only by the fact that Mexico city is most populated region of the country.  Thus, our future work might focus on experiments with three-dimensional substrates, which adequately reflect true landscape of Mexico, and on adding a historical perspective to our studies, particularly in choosing more ancient urban region for plasmodium inoculation.

Comparison of Physarum-made transport networks developed in United Kingdom~\cite{adamatzky_jones_2009} to the networks grown in Mexico brings interesting insights on the existing motorway structure in these countries. Physarum graphs, which built of only edges with highest probability of occurrence in experiments were minimum (for UK) and almost-minimum (for Mexico) spanning trees. Physarum graph is a sub-graph of UK motorways graph, while Physarum graph is a super-graph of Mexican highway graph. Does it mean that Mexican Federal Highway are redundant? Or do they simply reflect harsher geographical conditions which lead to a need of duplicating traffic routes? Further studies might bring answers to these questions.

\section*{Acknowledgement}
Genaro J. Mart{\'i}nez thanks DGAPA-UNAM and EPSRC for support. Authors express their gratitude to Mexican government and INEGI for providing the last census information and to Natalia Volkow Fern\'andez, and Gu{\'i}a Roji for permission to use additional road maps.


\begin{thebibliography}{99}


\bibitem{achenbach_1981}
Achenbach~F. and Weisenseel~M.~H.
 Ionic currents traverse the slime mould Physarum.
Cell Biol Int Rep. 1981 (5) 375--379.


\bibitem{adamatzky_2005}
Adamatzky A., De Lacy Costello B., Asai T. 
Reaction-Diffusion Computers, Elsevier, Amsterdam, 2005.

\bibitem{adamatzky_naturewissenschaften_2007}
Adamatzky A. 
Physarum machines: encapsulating reaction-diffusion to compute spanning tree. 
Naturwisseschaften 94 (2007) 975--980.


\bibitem{adamatzky_ppl_2007}
Adamatzky A. 
Physarum machine: implementation of a Kolmogorov-Uspensky machine on a biological substrate. 
Parallel Processing Letters 17 (2007) 455--467.

\bibitem{adamatzky_UC07}
Adamatzky~A. From reaction-diffusion to Physarum computing. Invited talk at Los Alamos Lab workshop 
``Unconventional Computing: Quo Vadis?'' (Santa Fe, NM, March 2007).


\bibitem{adamatzky_ppl_2009}
Adamatzky~A. Developing proximity graphs by \emph{Physarum polycephalum}: does the plasmodium 
follow the Toussaint hierarchy? Parallel Processing Letters (2009), in press.


\bibitem{adamatzky_bz_trees}
Adamatzky~A.
If BZ medium did spanning trees these would be the same trees as Physarum built.
Physics Letters A (2009), in press. 

\bibitem{adamatzky_hotice}
Adamatzky~A. Hot ice computer. Physics Lett A 374 (2009) 264--271.


\bibitem{adamatzky_jones_2009}
Adamatzky~A. and Jones~J.
Road planning with slime mould: If Physarum built motorways it would route M6/M74 through Newcastle 
Int J Bifurcaton and Chaos (2010), in print.
\url{http://arxiv.org/abs/0912.3967}

\bibitem{adamatzky_physarummachines}
Adamatzky~A. Physarum Machines: Making Computers from Slime Mould (World Scientific, 2010).


\bibitem{billiot_2010}
Billiot~J.~M., Corset~F., Fontenas~E.
Continuum percolation in the relative neighborhood graph.
\url{arXiv:1004.5292}


\bibitem{dorigo_2004}
Dorigo~M. and Stutzle~T.
Ant Colony Optimization
MIT Press, 2004.

\bibitem{gabriel_sokal_1969}
Gabriel~K.~R. and R. R. Sokal.
A new statistical approach to geographic variation analysis.  Systematic Zoology, 18 (1969) 259--278.


\bibitem{jaromczyk_toussaint_1992}
Jaromczyk~J.~W. and G.~T.~Toussaint, 
Relative neighborhood graphs and their relatives. Proc. IEEE 80 (1992) 1502--1517.


\bibitem{jarret_2006}
Jarrett~T.~C., Ashton~D.~J., Fricker~M., Johnson~N.~F.
Interplay between function and structure in complex networks
Phys. Rev. E 74 (2006) , 026116.

\bibitem{jones_transportnet_2008}
Jones~J.
The emergence and dynamical evolution of complex transport networks from simple low-level behaviours.
nt. Journal of Unconventional Computing, 6 (2010) 125--144.


\bibitem{kirkpatrick}
Kirkpatrick~D.~G. and Radke~J.~D.  
A framework for computational morphology. In G. Toussaint, editor, Computational Geometry (1985) 217--248. 



\bibitem{li_2004}
Li~X.-Y. 
Application of computation geometry in wireless networks. 
In: Cheng~X., Huang~X., Du~D.-Z. (Eds.)
Ad Hoc Wireless Networking (Kluwer Academic Publishers, 2004) 197--264.

\bibitem{nakagaki_yamada_1999}
Nakagaki~T. Yamada~H., Ueda~T.
Modulation of cellular rhythm and photoavoidance by
oscillatory irradiation in the Physarum plasmodium.
Biophysical Chemistry 82 (1999) 23--28.


\bibitem{nakagaki_2000}
Nakagaki T., Yamada H., Ueda T. 
Interaction between cell shape and contraction pattern
in the {\it Physarum plasmodium}, Biophysical Chemistry 84 (2000) 195--204.

\bibitem{nakagaki_2001}
Nakagaki T., 
Smart behavior of true slime mold in a labyrinth. 
Research in Microbiology 152 (2001) 767-–770.


\bibitem{nakagaki_2001a}
Nakagaki T., Yamada H., and Toth A.,
 Path finding by tube morphogenesis in an amoeboid organism. 
Biophysical Chemistry 92 (2001) 47–-52.

\bibitem{nakagaki_iima_2007}
Nakagaki~T., Iima~M., Ueda~T., Nishiura~y., Saigusa~T., Tero~A., Kobayashi~R., Showalter~K.
Minimum-risk path finding by an adaptive amoeba network.
Physical Review Letters 99 (2007) 068104.

\bibitem{matula_sokal_1984}
Matula~D.~W. and Sokal~R.~R. Properties of Gabriel graphs relevant to geographical variation research 
and the clustering of points in the same plane. Geographical Analysis 12 (1984) 205--222.

\bibitem{muhammad_2007}
Muhammad~R.~B.
A distributed graph algorithm for geometric routing in ad hoc wireless networks.
J Networks 2 (2007) 49--57. 

\bibitem{pointer_2005}
Pointer~G. Focus on People and Migration 2005. 
Chapter~3: The UK's major urban areas. 
(UK Statistics Authority, 2005)
\url{www.statistics.gov.uk}

\bibitem{reyes_2002}
Reyes~D.~R., Ghanem~M.~G., George~M. 
Glow discharge in micro fluidic chips for visible analog computing. 
Lab on a Chip 1 (2002) 113--116.


\bibitem{saigusa}
Saigusa~T., Tero~A., Nakagaki~T., Kuramoto~Y.
Amoebae anticipate periodic events.
Phys Rev Lett. 2008 (100) 018101. 


\bibitem{santi_2005}
Santi~P. Topology Control in Wireless Ad Hoc and Sensor Networks (Wiley, 2005).

\bibitem{schumann_adamatzky_2009}
Schumann~A. and Adamatzky~A. 
Physarum spatial logic. In: Proc. 
1th Int. Symp. on
Symbolic and Numeric Algorithms for Scientific Computing
(Timisoara, Romania, September 26-29, 2009).

\bibitem{shirakawa}
Shirakawa~T., Adamatzky~A., Gunji~Y.-P., Miyake~Y. 
On simultaneous construction of Voronoi diagram and Delaunay triangulation by Physarum polycephalum. 
Int. J. Bifurcation and Chaos (2009), in press.



\bibitem{song_2004}
Song~W.-Z., Wang~Y., Li~X.-Y.
Localized algorithms for energy efficient topology in wireless ad hoc networks.
In: Proc. MobiHoc 2004 (May 24–-26, 2004, Roppongi, Japan).

\bibitem{sridharan_2010}
Sridharan~M. and Ramasamy~A.~M.~S.
Gabriel graph of geomagnetic Sq variations.
Acta Geophysica (2010) \url{10.2478/s11600-010-0004-y}

\bibitem{tero_2006}
Tero~A., Kobayashi~R. and Nakagaki~T.,
\emph{Physarum} solver: A biologically inspired method of road-network navigation.  
Physica A 363 (2006) 115--119.


\bibitem{toussaint_1980}
Toussaint G.~T., The relative neighborhood graph of a finite planar set, 
Pattern Recognition 12 (1980) 261--268.

\bibitem{tsuda_2004}
Tsuda~S., Aono~M., Gunji~Y.-P. 
Robust and emergent Physarum logical-computing. Biosystems 73 (2004) 45--55.

\bibitem{tsuda_2007}
Tsuda~S, Zauner~K.-P., Gunji Y.-P. 
Robot control with bio-logical cells. BioSystems 87 (2007) 215--223.


\bibitem{wan_2007}
Wan~P.-J., Yi~C.-W.
On the longest edge of Gabriel Graphs in wireless ad hoc networks.
IEEE Trans. on Parallel and Distributed Systems 18 (2007) 111--125.

\bibitem{watanabe_2005}
Watanabe~D. 
A study on analyzing the road network pattern using proximity graphs.
J  of the City Planning Institute of Japan 40 (2005) 133--138.

\bibitem{watanabe_2008}
Watanabe~D.
Evaluating the configuration and the travel efficiency on proximity graphs
as transportation networks. Forma 23 (2008) 81-–87. 

\end{thebibliography}
\end{document}